\documentclass[aps,pra,floatfix,superscriptaddress,showpacs,twocolumn]{revtex4}
\usepackage{amsmath}
\usepackage{graphicx}

\begin{document}
\title{Optical Magnetometry}
\author{Dmitry Budker}
\email{budker@berkeley.edu} \affiliation{Department of Physics,
University of California, Berkeley, CA 94720-7300}
\affiliation{Nuclear Science Division, Lawrence Berkeley National
Laboratory, Berkeley CA 94720}
\author{Michael Romalis}\email{romalis@princeton.edu}
\affiliation{Department of Physics, Princeton University, Princeton,
NJ 08544 }

\date{\today}
\begin{abstract}
Some of the most sensitive methods of measuring magnetic fields
utilize interactions of resonant light with atomic vapor. Recent
developments in this vibrant field are improving magnetometers in
many traditional areas such as measurement of geomagnetic anomalies
and magnetic fields in space, and are opening the door to new ones,
including, dynamical measurements of bio-magnetic fields, detection
of nuclear magnetic resonance (NMR), magnetic-resonance imaging
(MRI), inertial-rotation sensing, magnetic microscopy with cold
atoms, and tests of fundamental symmetries of Nature.
\end{abstract}


\maketitle

\section{Measuring magnetic fields with atoms and light}

It has been nearly half a century since the techniques of measuring
magnetic fields using optical pumping and probing of alkali atoms
were realized in the pioneering work of Dehmelt \cite{Deh57}, Bell
and Bloom \cite{Bel57,Bel61a} and further developed by
Cohen-Tannoudji, Dupon-Roc, and co-workers, see, for example, Refs.
\cite{Dup69,Dup70}. 
The general idea of the method is
that light which is near-resonant with an optical transition creates
long-lived 
orientation and/or higher order moments in the atomic ground state
that subsequently undergo Larmor spin precession in the magnetic
field. This precession modifies the optical absorptive and
dispersive properties of the atoms
and this modification is detected by measuring the light transmitted
through the atomic medium. Recent reviews of resonant magneto-optics
have been given in Refs. \cite{Bud2002RMP,Ale2005}.

The fields of resonant magneto-optics and atomic magnetometry have
been experiencing a new boom driven by technological developments,
specifically by the advent of reliable, small, inexpensive, and
easily tunable diode lasers on the one hand, and by the refinement
of the techniques of producing dense atomic vapors with long (in
some cases $\sim 1\ $s) ground-state relaxation times on the other.
These technical advances have enabled atomic magnetometers to
achieve sensitivities rivaling
\cite{Ale2004,Gil2001,Wei2005,Bud2000Sens} and even surpassing
\cite{Kom2003} that of most Superconducting Quantum Interference
Device (SQUID) based magnetometers that have dominated the field of
sensitive magnetometers for a number of years \cite{Cla96}. Atomic
magnetometers have the intrinsic advantage of not requiring
cryogenic cooling, and offer a significant potential for
miniaturization. In contrast to SQUIDs that measure magnetic flux
through a pick-up loop, atomic magnetometers measure magnetic field
directly. Atomic magnetometers can be configured so that their
output is related to the magnitude of the magnetic field through
fundamental physical constants, so that no calibration is required.

Presently, the most sensitive atomic optical magnetometer is the
spin-exchange-relaxation-free (SERF) magnetometer whose demonstrated
sensitivity exceeds $10^{-15}\ $T/$\sqrt{\rm{Hz}}$, with projected
fundamental limits below $10^{-17}\ $T/$\sqrt{\rm{Hz}}$
\cite{Kom2003}. SERF magnetometers also offer a possibility of
spatially-resolved measurements with millimeter resolution.

The present-day interest in optical magnetometers is driven by
numerous and diverse applications, a partial list of which includes
tests of the fundamental symmetries of Nature, search for man-made
and natural magnetic anomalies, investigation of the dynamics of the
geomagnetic fields (including attempts at earthquake prediction),
material science and investigation of magnetic properties of rocks,
detection of magnetic microparticles at ultra-low concentrations,
detection of signals in nuclear magnetic resonance (NMR) and
magnetic-resonance imaging (MRI), direct detection of magnetic
fields from the heart and the brain, magnetic microscopy, and
measuring magnetic fields in space.

In this paper, we outline the basic principles and fundamental
limits of the sensitivity of optical atomic magnetometers, and
discuss several specific applications.

\section{General features and limits of sensitivity of optical
magnetometers} 
\begin{figure}[h]
\includegraphics[width=2.75 in]{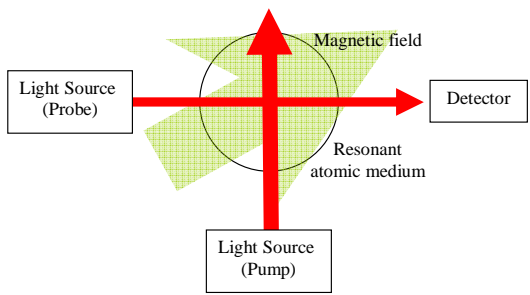}
\caption{A general schematic of an all-optical atomic magnetometer.
Pump light polarizes the atoms, atomic polarization evolves in the
magnetic field, and the resultant state of the atoms' polarization
is detected by measuring transmission or polarization rotation of
the probe light.}\label{Fig_General_Sch}
\end{figure}

A general schematic of an optical atomic magnetometer
is shown in Fig. \ref{Fig_General_Sch}. In many magnetometers, the
resonant medium is a vapor of alkali atoms (Rb, Cs, or K) contained
in a glass bulb. Because atomic polarization is generally destroyed
when atoms collide with the walls of the bulb, cells filled with
buffer gas are commonly used. The gas ensures that the atoms
optically polarized in the central part of the cell take a long time
to diffuse to the walls. Another technique for reducing wall
relaxation is application of a non-relaxing coating, typically,
paraffin, on the cell walls (see below).
As mentioned above, the light sources of choice today are diode
lasers; however, discharge lamps -- the original light sources for
atomic magnetometers -- are still used in most commercial atomic
magnetometers, and can achieve sensitivity comparable to lasers in
some research applications \cite{Gro2005}. Figure
\ref{Fig_General_Sch} shows separate light sources for pumping and
probing with orthogonal light beams. There exist many magnetometer
configurations that differ by relative direction and spectral tuning
of the pump and probe light. In some schemes, pumping and probing is
performed by the same beam. The two most common detection modes are
monitoring the intensity and polarization of the transmitted probe
light. The latter method has certain intrinsic advantages, such as
its ability to detect very small polarization-rotation angles, and a
reduced sensitivity to the laser-intensity noise. Shown in Fig.
\ref{Fig_General_Sch} is an all-optical magnetometer; no other
electromagnetic fields are applied to the atoms apart from the
magnetic field being measured and the pump and probe light. Some
magnetometers require additional means for excitation of spin
precession.  A weak  magnetic field oscillating at the Larmor
frequency is commonly used for this purpose \cite{Bel57,Wei2005}.
Other techniques include application of microwave fields
\cite{Ver2000} and all-optical excitation using various types of
modulation of the light beams: intensity, frequency or polarization
\cite{Bel61a,Ale2005}.

Quantum mechanics sets fundamental limits on the best sensitivity
that can be achieved in a magnetic-field measurement using atoms.
One such limit is associated with projection noise resulting from
the fact that if an atom is polarized in a particular direction, a
measurement of the angular-momentum projection $m$ on an orthogonal
direction yields a random result (+1/2 or -1/2 in the simplest case
of angular momentum $F=1/2$). Ignoring factors of order unity that
depend on particulars of the system (for example, the total value of
the angular momentum $F$, and the relative contributions of
different Zeeman sublevels), the sensitivity of a magnetic-field
measurement performed for a time $T$ with an ensemble of $N$ atoms
with coherence time $\tau$ is
\begin{equation}
\delta{B}\simeq\frac{1}{g\mu_B}\frac{\hbar}{\sqrt{N\tau{T}}}\,,
\label{Eqn:SQL}
\end{equation}
where $\mu_B$ is the Bohr magneton, $g$ is the ground-state
Land\'{e} factor, and $\hbar$ is the Planck's constant. Equation
\eqref{Eqn:SQL} is derived by noticing that a measurement with a
single atom with a duration of $\tau$ produces an uncertainty in the
Larmor precession angle on the order of $1$ rad. With $N$ atoms,
this is improved by $\sqrt{N}$, and repeating the measurement
multiple times gains another factor of $\sqrt{T/\tau}$.
%

Recently, a possibility of overcoming the projection noise in
magnetometry using spin-squeezing techniques -- where quantum states
are prepared with unequal distribution of uncertainty between
conjugate observables, in this case, the projections of the angular
momentum on two orthogonal directions -- was discussed and
demonstrated (see Ref. \cite{Ger2005} and references therein).
Unfortunately, an improvement in sensitivity using spin squeezing
appears possible only on a time scale significantly shorter than the
spin-relaxation time \cite{Auz2005}.

In optical magnetometry, in addition to the atomic projection noise,
there is also photon shot noise. For example, if the measured
quantity is the rotation angle $\varphi$ of the light polarization,
the shot noise is
\begin{equation}
    \delta\varphi_s\simeq\frac{1}{2\sqrt{\dot{N}_{ph}T}}~.
    \label{eqn:PolarSensitivity}
\end{equation}
Here $\dot{N}_{ph}$ is the probe-photon flux (in photons/s) after
the atomic sample. Depending on the details of a particular
measurement, either the spin noise (\ref{Eqn:SQL}) or the photon
noise (\ref{eqn:PolarSensitivity}) may dominate. However, if a
measurement is optimized for statistical sensitivity, the two
contributions to the noise are generally found to be comparable
\cite{Bud2000Sens,Sav2005RF}.

Another potential source of noise in atomic magnetometers is the AC
Stark shift caused by the probe and/or pump laser, which generates a
fictitious magnetic field proportional to the degree of circular
polarization of the light \cite{Hap67b}.
Even in the absence of technical sources of intensity or
polarization fluctuations, quantum fluctuations generate noise of
the fictitious magnetic field \cite{Fle2000}.
However, the noise due to AC Stark shifts can, in principle, be
eliminated by choosing a laser frequency where it crosses zero
\cite{NovAc2001}
or a geometry where the fictitious field is orthogonal to the
measured magnetic field. Nevertheless, in practical implementations
of magnetometers, light shifts due to drifts of laser properties are
often a significant concern.

The  ultimate sensitivity of atomic magnetometers is given by the
product of three quantities in  Eq. \eqref{Eqn:SQL}, the magnetic
moment of the atoms ($g\mu_B$), the square root of the number of
atoms involved in the
measurement, and the square root of the spin-relaxation time. 
Consequently, to
improve the sensitivity of a magnetometer, the number 
of atoms in the system and their spin-relaxation time should be
maximized.

There are several mechanisms that limit spin-relaxation time, one of
the most important being depolarization caused by collisions with
the cell walls that enclose the atomic vapor. Surface relaxation can
be reduced by using a coating that has a low adsorption energy for
atoms, so they spend less time bound to the surface of the cell.
Among such coatings, paraffin and other materials with long chains
of hydrocarbons were found to work well with alkali metals
\cite{Rob58}. In a seminal study \cite{BouBro66} Bouchiat and
Brossel demonstrated that spin relaxation on paraffin is caused by
two effects of comparable size, magnetic dipolar interaction and
spin-rotation coupling. Magnetic dipolar relaxation is dominated by
interaction between the magnetic moment of the electron and magnetic
moments of protons in the coating. They showed that replacing
hydrogen with deuterium, which has about three times smaller nuclear
magnetic moment, reduces this type of relaxation.
Despite work by several groups over the years, surface coating is
still a rather laborious process with some degree of ``black magic''
that does not always yield reproducible results. While collisions
with bare glass are generally completely depolarizing for alkali
atoms, high-quality coatings can allow more than 10,000 bounces
before depolarization, which also implies that even if a small
fraction of the surface has defects, this
will ruin the performance of the coating.


Another way to improve magnetometer sensitivity is to increase the
density of alkali-metal atoms, as was studied, for example, in Ref.
\cite{Sau2000}. This typically requires increasing the temperature
of the cell, although alternative approaches using light-induced
desorption have been investigated \cite{AleLIAD}. Since paraffin
coatings do not work at temperatures higher than $\approx 80^0\ $C,
high-density magnetometers usually use a buffer gas to slow down the
diffusion of alkali-metal atoms to cell walls.  Slow atomic
diffusion combined with spatially resolved optical detection also
allows many independent measurements of the magnetic field inside
one cell \cite{Kom2003}. As the density is increased, at some point,
the spin relaxation time becomes dominated by collisions between
alkali-metal atoms  and the product $N \tau$ approaches a constant,
so that shot-noise sensitivity no longer increases with density.
Thus, relaxation due to collisions between alkali-metal atoms
represents a fundamental obstacle to improvement in sensitivity for
a given cell volume.

Collisions between alkali-metal atoms are dominated by the
spin-exchange process in which the electron spins of the colliding
atoms rotate with respect to their combined spin, which is conserved
in the collision. Even though such collisions conserve the total
spin, they can lead to loss of spin coherence. All alkali atoms have
non-zero nuclear spin, $I$, and their ground states are split into
two hyperfine-structure components, characterized by the total
angular momentum $F = I \pm 1/2$. The direction of magnetic
precession, determined by the relative orientation of the electron
spin with respect to the total angular momentum, is opposite for the
two hyperfine states. Thus, in the presence of a magnetic field, the
spin-exchange collisions that randomly transfer atoms between the
two hyperfine states normally lead to spin-relaxation, as atomic
angular momenta acquire random angles with respect to each other.

As was first realized by Happer \cite{Hap73,Hap77}, it is possible
to suppress the effects of spin-exchange relaxation by {\it
increasing} the rate of spin-exchange collisions until it exceeds
the Larmor precession frequency. The effect is quite similar to
Dicke narrowing \cite{Dic53} in microwave and optical spectroscopy
or motional narrowing in NMR. In the rapid spin-exchange regime, we
may no longer speak of magnetic precession of atoms in an individual
hyperfine state. Instead, each atom experiences an average
precession in the same direction as would a free atom in the $F = I
+ 1/2$ state. This is because atoms, while being redistributed among
the sublevels, spend more time in this state, which has a higher
statistical weight and higher electron-spin polarization. As a
result, in a weak external magnetic field, the average angular
momentum of the atomic vapor precesses without spin-exchange
relaxation, although at a rate that is slower than the precession
rate for a free atom. The slowing down of the spin precession rate
depends on the distribution of atoms among magnetic sublevels and is
thus a sensitive function of the optical pumping process
\cite{Sav2005SE}. This dependence on local optical pumping
conditions may lead to non-uniformity of the spin precession
frequency over the volume of the cell and broadening of the
resonance signal. The problem can be avoided by operating the
magnetometer near zero magnetic field.

In this regime the Zeeman resonance linewidth is not broadened at
all by spin-exchange collisions and the alkali metal density can be
increased to about 10$^{14}$ cm$^{-3}$, four orders of magnitude
higher than in traditional atomic magnetometers. Eventually the
relaxation time becomes limited by ``spin-destruction'' collisions
between alkali-metal atoms that do not conserve the total spin of
the colliding pair. Several mechanisms of comparable importance have
been identified for such relaxation \cite{Eri2000}, but some aspects
of this process remain poorly understood \cite{Kad98}. The measured
spin-destruction cross-sections are smaller for lighter alkali-metal
atoms and result in a fundamental limit on the sensitivity of a K
magnetometer of about 10$^{-17}V^{-1/2}\ $T/Hz$^{1/2}$
\cite{All2002}, where $V$ is the active volume of the sensor in
cm$^3$.
%

The spin-exchange-relaxation-free regime can be achieved only in a
magnetic field range of less than about 10 nT. At higher magnetic
fields such as the Earth's field, it is possible to use a single
SERF magnetometer as a 3-axis null-detector with external feedback
\cite{Sel2004}.
The possibility of reducing spin-exchange relaxation in a finite
magnetic field has also been explored \cite{App99}. The idea is to
pump most atoms into a stretched state with $m=I+1/2$ where they
cannot undergo spin-exchange collisions due to conservation of
angular momentum. This technique works at any magnetic field, but it
cannot completely eliminate spin-exchange relaxation, since a high
optical pumping rate required to put atoms into the stretched spin
state in the presence of other relaxation processes also contributes
to the resonance linewidth. For an optimal pumping rate the minimum
resonance linewidth is given by the geometric mean of the
spin-exchange and spin-destruction rates \cite{Jau2004,Sav2005RF}.
This method for reduction of spin-exchange broadening has been used
for narrowing of a microwave resonance in an alkali-metal atomic
clock \cite{Jau2004}
and to improve the sensitivity of a resonant magnetometer for
detection of very weak RF fields \cite{Sav2005RF}. However, this
technique does not significantly improve the sensitivity for
measurements of static fields because excitation of large spin
coherences, necessary to obtain an optimal signal/noise ratio,
removes the atoms from the stretched spin state \cite{Smu2006}.

\begin{figure}
    \includegraphics[scale=0.8]{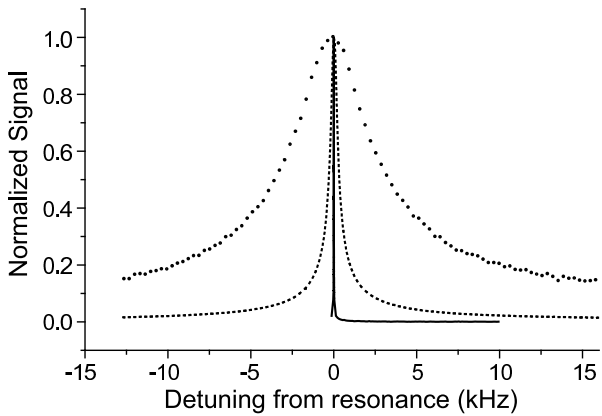}
    \caption{Comparison of Zeeman resonances for
     different modes of operation in a
    potassium vapor with density $n=7 \times 10^{13}$ cm$^{-3}$.
Points:  Spin-exchange broadened resonance with a full width at half
maximum of 3 kHz  observed in a magnetic field of
$10\ \mu$T when spin polarization is low. Dashed line: Resonance is
narrowed to a full width of 350 Hz at the same magnetic field by
pumping a large fraction of atoms into a stretched state parallel to
the magnetic field. Solid line: At a field of 5 nT, the resonance
width is 2 Hz due to complete elimination of spin-exchange
broadening in the the rapid spin-exchange regime. (Adapted from Ref.
\cite{Sav2005RF}.)} \label{fig:spinex}
\end{figure}

In Fig. \ref{fig:spinex} we show the Zeeman resonance curves
obtained in a dense alkali-metal vapor in three different regimes.
Spin-exchange broadening for low spin polarization is reduced by a
factor of 10 by pumping most atoms into a stretched spin state. The
linewidth can be further reduced by a factor of more than 100 in a
very low magnetic field, where spin-exchange relaxation is
completely eliminated.

\section{Additional characteristics of a magnetometer}
Apart from sensitivity, there are many other characteristics of a
magnetometer that are important for specific applications. We have
already mentioned that some of the most sensitive magnetometers
operate best at relatively small magnetic fields, i.e., they have a
limited \emph{dynamic range}. An important benchmark for the
magnetometer's dynamic range is the geomagnetic field $\sim 50\
\mu$T of interest in many applications. While traditional rf-optical
double resonance magnetometers have operated in this range from
their inception (see, for example, Ref. \cite{Blo62}),
diode-laser-based all-optical magnetometers for the geophysical
range have been also developed recently (Refs.
\cite{Sta2001,And2003,Ali2005,Aco2006}, for example). At geomagnetic
fields, sensitive atomic magnetometers have to contend with the
complications arising from the nonlinear Zeeman effect caused by the
Breit-Rabi mixing of hyperfine energy levels.  The nonlinear Zeeman
effect leads to a splitting of the magnetic resonance into multiple
lines (see, for example, Ref. \cite{Aco2006}). Several approaches
have been proposed to alleviate the adverse effects of the nonlinear
Zeeman effect in all-optical magnetometers, including synchronous
optical pumping at the quantum revival frequency given by the
quadratic correction to the Zeeman energies (Ref. \cite{Sel2006} and
references therein), and selective excitation and detection of
coherences (that correspond to high-order polarization multipoles)
between ``stretched'' Zeeman sublevels unaffected by nonlinear
Zeeman effect \cite{Ale97,Yas2003Select,Pus2006pp}.

Another important property of a magnetometer is whether it is
\emph{scalar} or \emph{vector}, i.e., whether it measures the total
magnitude, or specific cartesian component(s) of the magnetic field.
While knowing all three vector components of a field provides a more
complete information about the field, a truly scalar sensor has an
advantage in that the device is insensitive to the orientation of
the sensor with respect to the field, which is important for
operation on a mobile platform.

Many magnetometer applications require operation at frequencies
lower than 1 Hz where $1/f$ noise is often dominant. Atomic
magnetometers have an intrinsic advantage over other types of
sensors in this regard because they use a sensing element with very
simple structure that does not generate intrinsic $1/f$ noise
usually observed in solid-state systems with many nearly degenerate
energy states \cite{Weissman88}.
In practical situations, significant $1/f$ noise may arise due to
external elements, such as laser fluctuations caused by air
currents. Such noise may be reduced using spin modulation techniques
\cite{Li2006}.

Atomic magnetometers operating in a finite field naturally tend to
be of the scalar type as they rely on the resonance between an rf
field or a modulated light field with Zeeman-split energy
eigenstates of an atom. However, there is a standard technique (see,
for example, Refs. \cite{Gra2001,Ale2004VAR,Sel2004}) for converting
a scalar sensor into a vector one that relies on the fact that if a
small bias field is applied to the sensor in a certain direction in
addition to the field to be measured, then the change in the overall
field magnitude is linear in the projection of the bias field on the
main field, and is only quadratic (and generally negligible) in the
projection on the orthogonal plane. Thus, applying three orthogonal
bias fields consecutively, and performing three measurements of the
overall magnetic-field magnitude, one reconstructs the overall field
vector. In practice, it may be convenient to apply all three bias
fields simultaneously and modulate them at different frequencies.
Synchronous detection of the magnetometer output at a corresponding
frequency yields the value of the cartesian component of the
magnetic field being measured.

For practical operation at finite fields, atomic magnetometers
require a feedback loop to keep the frequency of the excitation
locked to resonance as the magnetic field is changing. One approach
is to use a phase-sensitive detector with an external feedback loop
and a voltage-controlled oscillator. Another approach, which is
often simpler, is a \emph{self-oscillating} magnetometer  that uses
the measured spin-precession signal to directly generate the
rf-field in a positive feedback loop \cite{Blo62}. All-optical
self-oscillating atomic magnetometers have been demonstrated
recently, utilizing transitions between hyperfine \cite{Mat2005} and
Zeeman sublevels \cite{Sch2005,Hig2006}.

An important characteristic of a magnetometer is how fast the device
responds to a change in the magnetic field. The time response of a
passive atomic magnetometer to a small variation in the magnetic
field is usually equivalent to a first-order low-pass filter with a
time constant $\tau$. Hence the natural bandwidth of such a
magnetometer is equal to $(2 \pi \tau)^{-1}$ Hz. If it is desirable
to make measurements on a time scale $T<\tau$, one can adjust the
operating parameters of the magnetometer, such as the probe beam
intensity, so $\tau=T$. If the number of atoms $N$ is fixed, then
according to Eq. \eqref{Eqn:SQL}, one loses in sensitivity as
$T^{-1}$ for short times. However, if the number of alkali atoms can
be increased and the spin-coherence time is limited by collisions
between alkali atoms, then the sensitivity decreases only as
$T^{-1/2}$. With fixed number of atoms and light power, it is
possible to increase the bandwidth of a magnetometer with external
feedback by using a large gain in the feedback loop. However, if the
bandwidth is increased by a factor $K$ over the natural bandwidth,
the magnetometer output noise also increases by the same factor $K$
\cite{Bec2005}.

It is also interesting to consider the response of a magnetometer to
an instantaneous change in the magnetic field. Since Larmor
precession has no inertia, a magnetometer based on such precession
responds instantaneously to a change in the field. Our knowledge of
the new value of the frequency will be at first very uncertain, but
will improve with time as $T^{3/2}$ (the best scaling for the
uncertainty of a single-tone-frequency determination from a noisy
signal \cite{Rif74}).
This is discussed in the context of a practical self-oscillating
magnetometer in Ref. \cite{Hig2006}, where the effect of additional
noise sources such as photodetector and amplifier noise are also
considered.

In the case of portable and space-borne magnetometers, important
characteristics include ``heading errors'' -- the dependence of the
reading of the magnetometer on the orientation of the sensor with
respect to the field being measured, as well as the existence of
``dead zones,'' i.e., spatial orientations where the magnetometer
loses its sensitivity. Other parameters of importance include size
and power consumption of the sensor system.
A recent trend is utilization of vertical-cavity surface-emitting
lasers (VCSEL) as light sources which provide on the order of a
milliwatt of light resonant with the D-lines of rubidium and cesium,
do not require an external cavity, and consume only a few milliwatts
of power. Miniaturization of the vapor cells to millimeter scales
can be done using more or less standard techniques (see, for
example, Ref. \cite{Bal2006} for a description of a prototype
optical-rotation magnetometer using a 3-mm diameter paraffin-coated
Cs cell). Another approach, particularly appealing for future mass
production of miniaturized low-cost magnetometers, is manufacturing
of an integrated sensor package incorporating a VCSEL laser, an
alkali-vapor cell, optics, and a detector using the wafer production
techniques well developed by the semi-conductor industry. The first
magnetometers based on this approach with a grain-of-rice sized
integrated sensor have been recently constructed
\cite{Sch2004,Kna2006}, demonstrating a sensitivity of $50\
$pT/$\sqrt{\rm{Hz}}$, with anticipated improvement by several orders
of magnitude with further optimization.

A favorable feature of magnetometers with a small vapor cell is
their reduced sensitivity to magnetic-field gradients that can lead
to additional spin relaxation, line broadening, and performance
degradation. Magnetometers utilizing buffer-gas free anti-relaxation
coated cells are also less sensitive to small field gradients,
because each atom samples the volume of the cell during its many
bounces between the walls in the course of a relaxation time. This
leads to a significant averaging of the magnetic-field
inhomogeneities. A systematic study of the effects of the gradients
on the Rb ground-state spin relaxation in a coated cell is presented
in Ref. \cite{Pus2006grad} along with a survey of earlier work.

\section{Applications}

\subsection{Biological magnetic fields}
Detection of magnetic fields of biological origin allows
non-invasive studies of the time dependence and spatial distribution
of biocurrents. Most biological magnetic field studies have focused
on detection of the fields from the heart and the brain.
Measurements of the magnetic fields generated by the heart
(magnetocardiography) provide richer diagnostic information about
heart function than traditional electrocardiography and do not
require placing electrical contacts on the patient \cite{Fen2005}.
Useful diagnostic information is obtained by measuring the spatial
distribution of the magnetic field during different parts of the
heart cycle. Most magnetocardiography studies have been performed in
magnetically shielded rooms, but they are considered too expensive
for clinical application, and widespread use of magnetocardiography
requires development of relatively low-cost sensors that can operate
in unshielded environment. Measuring magnetic fields generated by
the brain (magnetoencephalography) has been used extensively for
functional brain studies \cite{Ham93}.
Magnetic fields associated with a particular sensory input, such as
auditory, visual or tactile stimulation, are recorded by averaging
the signals over many presentations of the same stimulus. Detailed
measurements of the spatial distribution of the magnetic field
around the head allows one to identify regions of the brain that
become active during processing of the sensory input. However,
spatial localization is complicated by the fact that the inverse
problem of finding the current distribution responsible for a
particular magnetic field distribution does not have a unique
solution. Additional information, such as MRI data and sophisticated
numerical algorithms are used for spatial localization.
Magnetoencephalography also finds increasing use in clinical
diagnostic applications, for example for treatment of epilepsy
\cite{Pap2005}.

\begin{figure}
    \includegraphics[scale=0.8]{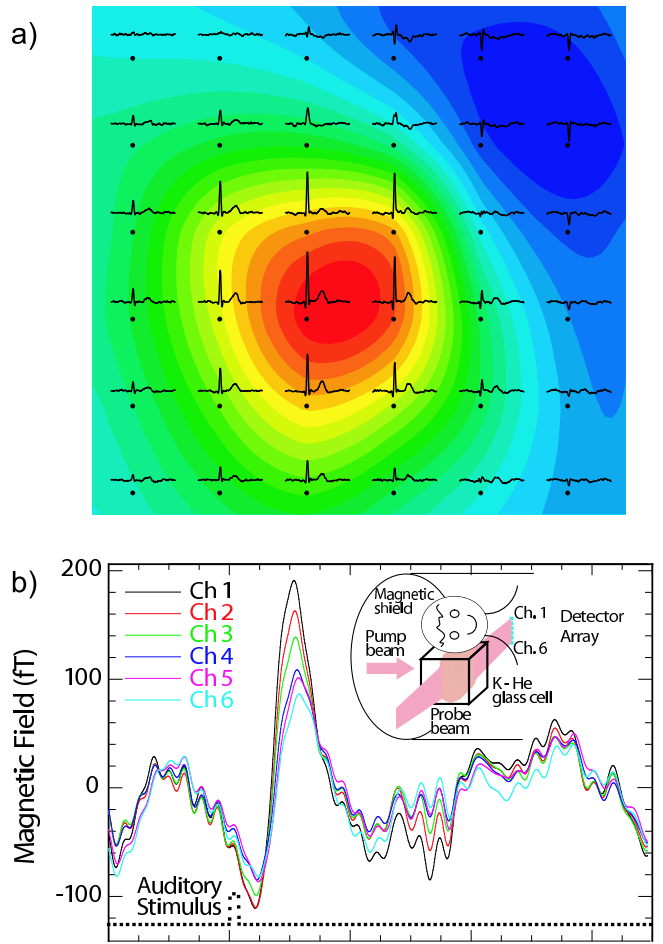}
    \caption{Examples of biological magnetic fields recorded with atomic
    magnetometers. a) Time traces (averaged over 100 heartbeats) of the out-of-chest component of the magnetic field from a human heart recorded on 36 grid points spaced by 4 cm.
    The underlying map is a snapshot of the field distribution at the moment of strongest magnetic activity (R-peak). The scale of the map ranges
    from -30 pT (blue) to +60 pT (red). Figure
    courtesy Prof. A. Weis.
    b) Magnetic fields recorded from a brain in response to an auditory stimulation by a
    series of short clicks (averaged over about 600 presentations). The  prominent feature at 100 ms after the
    stimulus is the evoked response in the auditory cortex, most clearly seen as a difference in the magnetic fields
    recorded by different channels. In contrast, ambient field drifts, such as seen before the stimulus,
    generate similar signals in all channels. Figure courtesy Dr. H. Xia.} \label{fig:biofield}
\end{figure}

The first measurements of biological fields with an atomic
magnetometer were performed in the 70s \cite{Liv77}
but this approach was not widely pursued and the majority of
biomagnetic applications relied on  SQUID magnetometers. Recent
progress in atomic magnetometry has again attracted interest in
their application for measurements of biological magnetic fields
with non-cryogenic sensors. Figure \ref{fig:biofield} shows examples
of magnetic fields from the heart and the brain detected with atomic
magnetometers. The cardio-magnetometer is based on optical-rf double
resonance and uses Cs atoms at $30^\circ$C, allowing it to be placed
close to a human body \cite{Bis2003}. Cardiomagnetic fields are
recorded sequentially on a grid of points above human chest.
Measurements of the brain magnetic field have been performed with a
potassium SERF magnetometer which operates at the vapor cell
temperature of $180^\circ$C and uses a multi-channel photodetector
to simultaneously record the spatial distribution of the magnetic
fields \cite{Xia2006}. Even though heating is required to maintain
the operating temperature of the cell positioned close to the
subject's head, it is technically easier and cheaper to do than
maintaining cryogenic temperature at the sensor, as required in the
case of SQUIDs.

\subsection{Fundamental applications}
Many fundamental interactions reduce at low energy to a spin
coupling similar to magnetic interaction since the spin is the only
vector available in the rest frame of the particle. Atomic
magnetometers are intrinsically sensitive probes of the spin
precession and hence play an important role in tests of fundamental
symmetries. An example of the magnetometers' versatile nature is a
Cs magnetometer constructed at Amherst. It was used to set a limit
on parity (P) and time-reversal invariance (T) violating electron
electric dipole moment (EDM) $d$ that generates $d \vec{E} \cdot
\vec{S}$ coupling \cite{Mur89}, on violation of Lorentz invariance
manifested as a spin coupling $\vec{b} \cdot \vec{S}$ to a
background vector field $\vec{b}$ \cite{Ber95}, and on
spin-dependent forces that can be mediated by axions \cite{You96}.

Magnetometers based on nuclear (rather than electron) spin
precession also play an important role in tests of fundamental
symmetries. While nuclei, such as $^3$He, have a magnetic moment
smaller than that of an electron by three orders of magnitude, they
also display much longer spin relaxation time $\tau$ and thus can be
competitive with electron-spin magnetometers in magnetic field
measurements \cite{Gil2003}.
In the searches for non-magnetic spin interactions, the smallness of
the nuclear magnetic moment is actually an advantage as it reduces
the sensitivity to spurious magnetic effects. Measurements of
nuclear spin precession of $^{199}$Hg have been used to set a limit
on the EDM of the mercury atom, which is also the tightest limit on
an EDM of any particle. This result constrains possible violation of
charge-parity (CP) symmetry due to supersymmetric particles
\cite{Rom2001PRL}. Comparison of $^{129}$Xe and $^3$He precession
sets the most stringent limit on the Lorentz-violating vector
coupling to a nuclear spin \cite{Bea2000,Bea2002}.

A number of new ideas for tests of fundamental symmetries using
atomic magnetometers are presently being explored, such as searches
for a permanent electric dipole moment using laser-cooled Cs atoms
in an optical lattice \cite{Chi2001}
or an atomic fountain \cite{Ami2006}, and the possibility of using
ultra-sensitive magnetometers for detecting P and T violating
magnetization in solid-state samples induced by an applied electric
field \cite{Lam2002} or an internal field of a ferroelectric (see
Ref. \cite{Bud2006CM} and references therein).
%

\subsection{Measuring magnetic fields in space}
Measurements of planetary and interplanetary magnetic fields have
been integral to space missions from their early days
\cite{Nes70,Acu97}. The on-board instruments designed for such
missions have been able to successfully meet various inherent design
challenges, including the necessity to have a very broad dynamic
range for the instrument, as the magnetic fields to be measured
could vary by many orders of magnitude between planetary fly-bys and
the craft's sojourn in interplanetary space. Other challenges
include stringent requirements on reliability, the ability to
withstand thermal and mechanical stresses associated with the launch
and varying conditions during the flight; limited weight and size;
power consumption, the ability to characterize and correct for
errors due to the craft spin, etc.

Most spacecraft measuring magnetic magnetic fields currently rely on
flux-gate magnetometers because of their small size and power
consumption. However, optically pumped $^4$He magnetometers are used
in most advanced space applications because of their relative
simplicity, reliability and high absolute accuracy. In a $^4$He
magnetometer, a weak electric discharge excites helium atoms to the
metastable 2$^3$S$_1$ state, and optical pumping with a $^4$He
discharge lamp is employed to polarize and detect spin precession of
atoms in the metastable state. Such instruments typically have
sensitivity on the order of $1-10\ $pT/$\sqrt{\rm{Hz}}$ and have
been successfully flown on Ulysses \cite{Bal88} and Cassini
\cite{Sou92,Dun99} missions, recently providing new information
about magnetospheres of Saturn \cite{Dou2005}
and its moon Enceladus \cite{Dou2006}.
Laser-pumped $^4$He magnetometers are presently being developed and
have demonstrated sensitivity of 200~fT$\sqrt{\rm{Hz}}$
\cite{Slo2002}
while their fundamental sensitivity limit  is estimated to be about
5~fT$\sqrt{\rm{Hz}}$ \cite{McG87}.

Deployment of more sensitive magnetometers in space will enable
magnetic field measurements in weak-field space environments, as in
the outer heliosphere and in the local interstellar medium. At
distances from the sun beyond about 80 astronomical units, the
strength of the ambient, nominally dc, magnetic field could be as
low as a few tens of picotesla (see Ref. \cite{Bur2005} and
references therein), near the limit of detectability of currently
used sensors. Fluctuations of magnetic fields in space are caused by
plasma process and have a bandwidth on the order of 1 Hz,
corresponding to typical electron cyclotron frequency in the outer
heliosphere. Atomic magnetometers are ideally suited for
measurements of such fields.

\subsection{Atomic magnetometers and nuclear magnetic resonance}
One of the rapidly growing applications of atomic magnetometers is
detection of NMR signals. NMR is usually detected with inductive rf
pick-up coils whose sensitivity drops at low frequency, precluding
many applications of low-field NMR. SQUID magnetometers have been
widely used for NMR detection at low frequencies \cite{Gre98}, but
they still require cryogenic cooling, negating one of the main
advantages of low-field NMR -- the absence of a cryogenic
superconducting magnet. Atomic magnetometers can be used in place of
SQUIDs to detect the magnetic field generated by the nuclear
magnetization \cite{Coh69b,Yas2004,Sav2005NMR}.

One of the promising applications is ``remote'' NMR, where spin
polarization, NMR-signal encoding and detection are performed
sequentially in different parts of the apparatus \cite{Mou2003}.
Such remote NMR \cite{Xu2006RSI} and MRI \cite{Xu2006IMAG} has
recently been demonstrated with an atomic magnetometer.

For both remote and in-situ NMR detection, having the NMR sample and
the magnetometer sensor spatially separated (as, for example, in
Refs. \cite{Xu2006RSI,Xu2006IMAG})  has experimental advantages,
such as an ability to apply an independent, relatively strong
magnetic field to the sample, which is not ``seen" by the sensor if
a proper geometrical arrangement is used. On the other hand, atomic
magnetometers can achieve an even higher sensitivity by taking
advantage of contact interactions between alkali-metal and nuclear
spins. These interactions have been particularly well studied for
noble gas atoms \cite{Sch89PRA} and can be described by a scalar
Fermi-contact interaction between the two spins. For heavy noble
gases, such as $^{129}$Xe, this interaction can enhance the magnetic
field produced by nuclear magnetization by a large factor on the
order of 600. Thus, by allowing noble gas atoms to interact directly
with the alkali-metal vapor, one can obtain high sensitivity to low
concentration of $^{129}$Xe spins \cite{Sav2005NMR}.

The contact interaction between nuclear and electron spins is also
useful for inertial rotation sensing. Nuclear spins make good
quantum gyroscopes because of their long spin coherence time, but
their magnetic moment causes significant precession in stray
magnetic fields \cite{Woo87}. An alkali-metal magnetometer can then
serve two purposes, to measure the ambient magnetic field and to
detect the inertial precession of nuclear spins with high
sensitivity. The requirement for magnetic field stability is rather
stringent. For example, a stray magnetic field of 1 fT would cause a
false spin precession rate for $^{3}$He of about 0.04 degrees/hour,
more than can be tolerated in a navigation-grade gyroscope. A SERF
magnetometer can achieve this level of magnetic sensitivity,
allowing it to be used for cancelation of stray magnetic fields and
enabling a competitive nuclear-spin rotation sensor \cite{Kor2005}.

Another new application of atomic magnetometers is for detection of
nuclear quadrupole resonance (NQR) signals. In crystalline materials
nuclei with a quadrupole moment are aligned by the electric field
gradient and can generate a weak rf signal following application of
a resonant rf pulse. The physics of this phenomenon can be
understood in the language of \emph{alignment-to-orientation
conversion} \cite{Bud2003NQR}, a process well studied in atomic
physics. For practical applications, it is particularly important
that NQR signals do not average out in materials with randomly
oriented crystallites such as powders. Resonance frequencies are
highly material-specific and typically range from 0.1 to 5 MHz.
Detection of NQR is a promising technique for identification of
explosives, since most explosive materials
contain the $^{14}$N nuclei, possessing a large quadrupole moment.
However, widespread use of NQR for this purpose has been limited due
to the weakness of the NQR signals, which are typically detected
with an rf coil only after substantial signal averaging
\cite{Gar2001}. Detection of weak radio-frequency signals requires
modification of usual atomic-magnetometer arrangements that are
designed for detection of quasi-DC magnetic fields. A tunable
magnetometer for detection of weak rf field can be realized by using
a bias field to tune the Zeeman energy splitting to the frequency of
the rf field \cite{Sav2005RF,Led2006}. Recently such magnetometer
has been built for operation at 423 kHz with a sensitivity of 0.24
fT/Hz$^{1/2}$ and bandwidth of 600 Hz and used to detect NQR signals
from ammonium nitride \cite{Lee2006}.
A detailed analysis of the fundamental limits on sensitivity of an
rf atomic magnetometer \cite{Sav2005RF} and an inductive pick-up
coil show that atomic magnetometer has higher sensitivity up to
frequencies of about 50 MHz \cite{Sav2006NMR_rf}.

\subsection{Magnetometery with cold atoms}
Recent breakthroughs in laser cooling and trapping have opened new
avenues for precision measurements using long-lived, near-stationary
collections of atoms. In particular, far-off-resonance optical traps
provide a benign environment for trapping atoms with negligible
photon scattering rates and storage lifetimes in excess of 300 s in
an ultra-high vacuum environment  \cite{OHa99}.
Thus, the spin-coherence time of laser-cooled and trapped atoms can
be substantially longer than the typical value of 1 s obtained in a
buffer gas or surface-coated cell. However, due to the small volume
of atom traps and limits on atomic number density from cold
collisions, the total number of trapped atoms is typically on the
order of $10^6-10^8$, many orders of magnitude smaller than
$10^{11}-10^{15}$ atoms contained in a cm-sized vapor cell. While
trapped atoms do not have the highest sensitivity to uniform
magnetic fields, they are particularly well suited for making field
measurements with high spatial resolution corresponding to the trap
size, where vapor cells do not work well because spin-relaxation
time quickly decreases with cell size. Such high resolution magnetic
microscopes find a range of applications, from studies of magnetic
domains \cite{Fre2001}
to imaging of currents on integrated circuits \cite{Cha2000}.

\begin{figure}
    \includegraphics[scale=0.45]{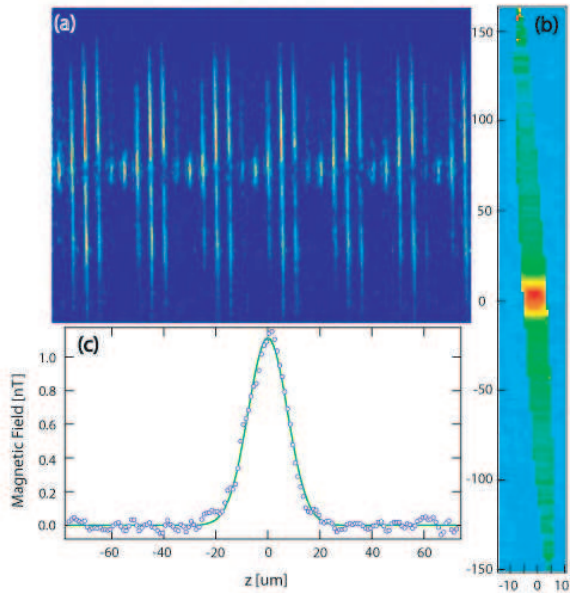}
    \caption{Detection of effective magnetic field by imaging of Larmor
precession in a Bose-Einstein condensate of $^{87}$Rb
\cite{Ven2006}. The condensate contained 1.4 million atoms,
optically trapped with trapping frequencies of 4.2 Hz $\times$ 155
Hz $\times$ 400 Hz. A uniform bias magnetic field of 16 $\mu$T is
applied, while a far-detuned circularly polarized laser beam is
focused onto the condensate with a waist of $7.6\,\mu$m to simulate
the effects of an additional localized magnetic field. Part (a)
shows a sequence of 32 individual normalized atomic-spin-sensitive
phase-contrast images taken after an evolution time of 50\,ms and
spaced in time by $100\,\mu$s.  The variation of the brightness of
the images with a period of $\approx$5 frames is due to Larmor
precession of the atomic spins.
Part (b) shows a pseudocolor map of the phase of Larmor precession
as a function of position in microns. The phase corresponds to the
effective magnetic field shown in Part (c) as a function of position
in the central portion of the condensate, demonstrating the high
resolution and precision of the technique. The gaussian phase shift
follows the intensity profile of the far-detuned laser beam, and the
solid line is a fit to this profile. Figure courtesy Drs. M.
Vengalattore and J. Higbie.} \label{fig:BECmag}
\end{figure}

Two magnetometery techniques have been recently demonstrated with Rb
Bose-Einstein condensates (BEC). The first involves holding the
condensate in a weak magnetic trap so the energy of interaction with
the magnetic field to be measured causes a perturbation of the
trapping potential and changes the local atomic density
\cite{Wil2005,Wil2006}.
With a typical chemical potential in a weak magnetic trap on the
order of 1 nK one gets a magnetic field sensitivity of about 1 nT.
The other technique \cite{Hig2005,Ven2006} is similar to
measurements done with hot atoms -- the BEC is held in an optical
dipole trap and spin precession is measured using phase-contrast
imaging with an off-resonant circularly-polarized probe beam. This
non-destructive imaging allows monitoring of spin precession, as
shown in Fig. \ref{fig:BECmag}. The magnetic field sensitivity
obtained with this method is 900 fT (in a single run with 250 ms
integration time and an integrating area of 100 square microns). In
both methods cooling atoms to quantum degeneracy improves
measurement sensitivity, but whether the coherence of the BEC plays
a direct role is presently an open question.






\section{Bright future}
Recent progress in atomic magnetometry techniques can be expected to
have a significant impact in four general areas. The first is the
development of robust laser-pumped atomic magnetometers, largely
driven by recent availability of electronically-tunable VCSEL and
distributed-feedback (DFB) semiconductor lasers. Such magnetometers
can replace existing discharge-lamp-pumped devices, providing higher
sensitivity in geological, military, and space applications. The
second area is the development of micro-fabricated mm-size
magnetometers, which will open entirely new applications of magnetic
monitoring in a wide range of environments. The third area is the
increasing use of atomic magnetometers as sensitive detectors for
weak signals, in NMR, biological applications, magnetic microscopy,
and inertial rotation sensing. The fourth area is the exploration of
the frontier of magnetic sensitivity, where atomic magnetometers are
already surpassing SQUID sensors and can be expected to reach
sensitivities significantly below $10^{-17}\ $T/Hz$^{1/2}$. This
will enable new measurements in materials science, improve the
precision of fundamental physics tests, and may lead to entirely
unexpected discoveries. The future of measuring magnetic fields with
atoms and light is indeed bright.

\section*{Acknowledgments}
This work is supported by DOD MURI grant \# N-00014-05-1-0406. We
are grateful to E. Alexandrov, M. Balabas, G. Bison, S. Bale, W.
Gawlik, J. Higbie, M. Ledbetter, I. M. Savukov, D. Stamper-Kurn, A.
Sushkov, M. Vengalattore, and A. Weis for providing valuable input
for this review.

\bibliography{NMObibl}

\begin{thebibliography}{107}
\expandafter\ifx\csname natexlab\endcsname\relax\def\natexlab#1{#1}\fi
\expandafter\ifx\csname bibnamefont\endcsname\relax
  \def\bibnamefont#1{#1}\fi
\expandafter\ifx\csname bibfnamefont\endcsname\relax
  \def\bibfnamefont#1{#1}\fi
\expandafter\ifx\csname citenamefont\endcsname\relax
  \def\citenamefont#1{#1}\fi
\expandafter\ifx\csname url\endcsname\relax
  \def\url#1{\texttt{#1}}\fi
\expandafter\ifx\csname urlprefix\endcsname\relax\def\urlprefix{URL }\fi
\providecommand{\bibinfo}[2]{#2}
\providecommand{\eprint}[2][]{\url{#2}}

\bibitem[{\citenamefont{Dehmelt}(1957)}]{Deh57}
\bibinfo{author}{\bibfnamefont{H.}~\bibnamefont{Dehmelt}},
  \bibinfo{journal}{Phys. Rev.} \textbf{\bibinfo{volume}{105}},
  \bibinfo{pages}{1924} (\bibinfo{year}{1957}).

\bibitem[{\citenamefont{Bell and Bloom}(1957)}]{Bel57}
\bibinfo{author}{\bibfnamefont{W.}~\bibnamefont{Bell}} \bibnamefont{and}
  \bibinfo{author}{\bibfnamefont{A.}~\bibnamefont{Bloom}},
  \bibinfo{journal}{Phys. Rev.} \textbf{\bibinfo{volume}{107}},
  \bibinfo{pages}{1559} (\bibinfo{year}{1957}).

\bibitem[{\citenamefont{Bell and Bloom}(1961)}]{Bel61a}
\bibinfo{author}{\bibfnamefont{W.}~\bibnamefont{Bell}} \bibnamefont{and}
  \bibinfo{author}{\bibfnamefont{A.}~\bibnamefont{Bloom}},
  \bibinfo{journal}{Phys. Rev. Lett.} \textbf{\bibinfo{volume}{6}},
  \bibinfo{pages}{280} (\bibinfo{year}{1961}).

\bibitem[{\citenamefont{Dupont-Roc et~al.}(1969)\citenamefont{Dupont-Roc,
  Haroche, and Cohen-Tannoudji}}]{Dup69}
\bibinfo{author}{\bibfnamefont{J.}~\bibnamefont{Dupont-Roc}},
  \bibinfo{author}{\bibfnamefont{S.}~\bibnamefont{Haroche}}, \bibnamefont{and}
  \bibinfo{author}{\bibfnamefont{C.}~\bibnamefont{Cohen-Tannoudji}},
  \bibinfo{journal}{Phys. Lett. A} \textbf{\bibinfo{volume}{28a}},
  \bibinfo{pages}{638} (\bibinfo{year}{1969}).

\bibitem[{\citenamefont{Dupont-Roc}(1970)}]{Dup70}
\bibinfo{author}{\bibfnamefont{J.}~\bibnamefont{Dupont-Roc}},
  \bibinfo{journal}{Rev. Phys. Appl.} \textbf{\bibinfo{volume}{5}},
  \bibinfo{pages}{853} (\bibinfo{year}{1970}).

\bibitem[{\citenamefont{Budker et~al.}(2002)\citenamefont{Budker, Gawlik,
  Kimball, Rochester, Yashchuk, and Weis}}]{Bud2002RMP}
\bibinfo{author}{\bibfnamefont{D.}~\bibnamefont{Budker}},
  \bibinfo{author}{\bibfnamefont{W.}~\bibnamefont{Gawlik}},
  \bibinfo{author}{\bibfnamefont{D.~F.} \bibnamefont{Kimball}},
  \bibinfo{author}{\bibfnamefont{S.~M.} \bibnamefont{Rochester}},
  \bibinfo{author}{\bibfnamefont{V.~V.} \bibnamefont{Yashchuk}},
  \bibnamefont{and} \bibinfo{author}{\bibfnamefont{A.}~\bibnamefont{Weis}},
  \bibinfo{journal}{Rev. Mod. Phys.} \textbf{\bibinfo{volume}{74}},
  \bibinfo{pages}{1153} (\bibinfo{year}{2002}).

\bibitem[{\citenamefont{Alexandrov et~al.}(2005)\citenamefont{Alexandrov,
  Auzinsh, Budker, Kimball, Rochester, and Yashchuk}}]{Ale2005}
\bibinfo{author}{\bibfnamefont{E.~B.} \bibnamefont{Alexandrov}},
  \bibinfo{author}{\bibfnamefont{M.}~\bibnamefont{Auzinsh}},
  \bibinfo{author}{\bibfnamefont{D.}~\bibnamefont{Budker}},
  \bibinfo{author}{\bibfnamefont{D.~F.} \bibnamefont{Kimball}},
  \bibinfo{author}{\bibfnamefont{S.~M.} \bibnamefont{Rochester}},
  \bibnamefont{and} \bibinfo{author}{\bibfnamefont{V.~V.}
  \bibnamefont{Yashchuk}}, \bibinfo{journal}{J. Opt. Soc. Am. B}
  \textbf{\bibinfo{volume}{22}}, \bibinfo{pages}{7} (\bibinfo{year}{2005}).

\bibitem[{\citenamefont{Aleksandrov et~al.}(2004)\citenamefont{Aleksandrov,
  Balabas, Vershovskii, and Pazgalev}}]{Ale2004}
\bibinfo{author}{\bibfnamefont{E.~B.} \bibnamefont{Aleksandrov}},
  \bibinfo{author}{\bibfnamefont{M.~V.} \bibnamefont{Balabas}},
  \bibinfo{author}{\bibfnamefont{A.~K.} \bibnamefont{Vershovskii}},
  \bibnamefont{and} \bibinfo{author}{\bibfnamefont{A.~S.}
  \bibnamefont{Pazgalev}}, \bibinfo{journal}{Technical Physics}
  \textbf{\bibinfo{volume}{49}}, \bibinfo{pages}{779} (\bibinfo{year}{2004}).

\bibitem[{\citenamefont{Gilles et~al.}(2001)\citenamefont{Gilles, Hamel, and
  Cheron}}]{Gil2001}
\bibinfo{author}{\bibfnamefont{H.}~\bibnamefont{Gilles}},
  \bibinfo{author}{\bibfnamefont{J.}~\bibnamefont{Hamel}}, \bibnamefont{and}
  \bibinfo{author}{\bibfnamefont{B.}~\bibnamefont{Cheron}},
  \bibinfo{journal}{Rev. Sci. Instrum.} \textbf{\bibinfo{volume}{72}},
  \bibinfo{pages}{2253} (\bibinfo{year}{2001}).

\bibitem[{\citenamefont{Weis and Wynands}(2005)}]{Wei2005}
\bibinfo{author}{\bibfnamefont{A.}~\bibnamefont{Weis}} \bibnamefont{and}
  \bibinfo{author}{\bibfnamefont{R.}~\bibnamefont{Wynands}},
  \bibinfo{journal}{Optics and Lasers in Engineering}
  \textbf{\bibinfo{volume}{43}}, \bibinfo{pages}{387} (\bibinfo{year}{2005}).

\bibitem[{\citenamefont{Budker et~al.}(2000)\citenamefont{Budker, Kimball,
  Rochester, Yashchuk, and Zolotorev}}]{Bud2000Sens}
\bibinfo{author}{\bibfnamefont{D.}~\bibnamefont{Budker}},
  \bibinfo{author}{\bibfnamefont{D.~F.} \bibnamefont{Kimball}},
  \bibinfo{author}{\bibfnamefont{S.~M.} \bibnamefont{Rochester}},
  \bibinfo{author}{\bibfnamefont{V.~V.} \bibnamefont{Yashchuk}},
  \bibnamefont{and}
  \bibinfo{author}{\bibfnamefont{M.}~\bibnamefont{Zolotorev}},
  \bibinfo{journal}{Phys. Rev. A} \textbf{\bibinfo{volume}{62}},
  \bibinfo{pages}{043403} (\bibinfo{year}{2000}).

\bibitem[{\citenamefont{Kominis et~al.}(2003)\citenamefont{Kominis, Kornack,
  Allred, and Romalis}}]{Kom2003}
\bibinfo{author}{\bibfnamefont{I.~K.} \bibnamefont{Kominis}},
  \bibinfo{author}{\bibfnamefont{T.~W.} \bibnamefont{Kornack}},
  \bibinfo{author}{\bibfnamefont{J.~C.} \bibnamefont{Allred}},
  \bibnamefont{and} \bibinfo{author}{\bibfnamefont{M.~V.}
  \bibnamefont{Romalis}}, \bibinfo{journal}{Nature}
  \textbf{\bibinfo{volume}{422}}, \bibinfo{pages}{596} (\bibinfo{year}{2003}).

\bibitem[{\citenamefont{Clarke}(1996)}]{Cla96}
\bibinfo{author}{\bibfnamefont{J.}~\bibnamefont{Clarke}}, in
  \emph{\bibinfo{booktitle}{SQUID Sensors: Fundamentals, Fabrication, and
  Applications}}, edited by
  \bibinfo{editor}{\bibfnamefont{H.}~\bibnamefont{Weinstock}}
  (\bibinfo{publisher}{Kluwer Academic}, \bibinfo{address}{The Netherlands},
  \bibinfo{year}{1996}), pp. \bibinfo{pages}{1--62}.

\bibitem[{\citenamefont{Groeger et~al.}(2005)\citenamefont{Groeger, Pazgalev,
  and Weis}}]{Gro2005}
\bibinfo{author}{\bibfnamefont{S.}~\bibnamefont{Groeger}},
  \bibinfo{author}{\bibfnamefont{A.~S.} \bibnamefont{Pazgalev}},
  \bibnamefont{and} \bibinfo{author}{\bibfnamefont{A.}~\bibnamefont{Weis}},
  \bibinfo{journal}{Appl. Phys. B.} \textbf{\bibinfo{volume}{80}},
  \bibinfo{pages}{645} (\bibinfo{year}{2005}).

\bibitem[{\citenamefont{Vershovskii et~al.}(2000)\citenamefont{Vershovskii,
  Pazgalev, and Aleksandrov}}]{Ver2000}
\bibinfo{author}{\bibfnamefont{A.~K.} \bibnamefont{Vershovskii}},
  \bibinfo{author}{\bibfnamefont{A.~S.} \bibnamefont{Pazgalev}},
  \bibnamefont{and} \bibinfo{author}{\bibfnamefont{E.~B.}
  \bibnamefont{Aleksandrov}}, \bibinfo{journal}{Technical Physics}
  \textbf{\bibinfo{volume}{45}}, \bibinfo{pages}{88} (\bibinfo{year}{2000}).

\bibitem[{\citenamefont{Geremia et~al.}(2005)\citenamefont{Geremia, Stockton,
  and Mabuchi}}]{Ger2005}
\bibinfo{author}{\bibfnamefont{G.~M.} \bibnamefont{Geremia}},
  \bibinfo{author}{\bibfnamefont{J.~K.} \bibnamefont{Stockton}},
  \bibnamefont{and} \bibinfo{author}{\bibfnamefont{H.}~\bibnamefont{Mabuchi}},
  \bibinfo{journal}{Phys. Rev. Lett.} \textbf{\bibinfo{volume}{94}},
  \bibinfo{pages}{203002} (\bibinfo{year}{2005}).

\bibitem[{\citenamefont{Auzinsh et~al.}(2004)\citenamefont{Auzinsh, Budker,
  Kimball, Rochester, Stalnaker, Sushkov, and Yashchuk}}]{Auz2005}
\bibinfo{author}{\bibfnamefont{M.}~\bibnamefont{Auzinsh}},
  \bibinfo{author}{\bibfnamefont{D.}~\bibnamefont{Budker}},
  \bibinfo{author}{\bibfnamefont{D.~F.} \bibnamefont{Kimball}},
  \bibinfo{author}{\bibfnamefont{S.}~\bibnamefont{Rochester}},
  \bibinfo{author}{\bibfnamefont{J.~E.} \bibnamefont{Stalnaker}},
  \bibinfo{author}{\bibfnamefont{A.~O.} \bibnamefont{Sushkov}},
  \bibnamefont{and} \bibinfo{author}{\bibfnamefont{V.}~\bibnamefont{Yashchuk}},
  \bibinfo{journal}{Phys. Rev. Lett.} \textbf{\bibinfo{volume}{93}},
  \bibinfo{pages}{173002} (\bibinfo{year}{2004}).

\bibitem[{\citenamefont{Savukov et~al.}(2005)\citenamefont{Savukov, Seltzer,
  Romalis, and Sauer}}]{Sav2005RF}
\bibinfo{author}{\bibfnamefont{I.~M.} \bibnamefont{Savukov}},
  \bibinfo{author}{\bibfnamefont{S.~J.} \bibnamefont{Seltzer}},
  \bibinfo{author}{\bibfnamefont{M.~V.} \bibnamefont{Romalis}},
  \bibnamefont{and} \bibinfo{author}{\bibfnamefont{K.~L.} \bibnamefont{Sauer}},
  \bibinfo{journal}{Phys. Rev. Lett.} \textbf{\bibinfo{volume}{95}},
  \bibinfo{pages}{0630041} (\bibinfo{year}{2005}).

\bibitem[{\citenamefont{Happer and Mathur}(1967)}]{Hap67b}
\bibinfo{author}{\bibfnamefont{W.}~\bibnamefont{Happer}} \bibnamefont{and}
  \bibinfo{author}{\bibfnamefont{B.}~\bibnamefont{Mathur}},
  \bibinfo{journal}{Phys. Rev.} \textbf{\bibinfo{volume}{163}},
  \bibinfo{pages}{12} (\bibinfo{year}{1967}).

\bibitem[{\citenamefont{Fleischhauer et~al.}(2000)\citenamefont{Fleischhauer,
  Matsko, and Scully}}]{Fle2000}
\bibinfo{author}{\bibfnamefont{M.}~\bibnamefont{Fleischhauer}},
  \bibinfo{author}{\bibfnamefont{A.~B.} \bibnamefont{Matsko}},
  \bibnamefont{and} \bibinfo{author}{\bibfnamefont{M.~O.}
  \bibnamefont{Scully}}, \bibinfo{journal}{Phys. Rev. A}
  \textbf{\bibinfo{volume}{62}}, \bibinfo{pages}{013808/1}
  (\bibinfo{year}{2000}).

\bibitem[{\citenamefont{Novikova et~al.}(2001)\citenamefont{Novikova, Matsko,
  Velichansky, Scully, and Welch}}]{NovAc2001}
\bibinfo{author}{\bibfnamefont{I.}~\bibnamefont{Novikova}},
  \bibinfo{author}{\bibfnamefont{A.~B.} \bibnamefont{Matsko}},
  \bibinfo{author}{\bibfnamefont{V.~L.} \bibnamefont{Velichansky}},
  \bibinfo{author}{\bibfnamefont{M.~O.} \bibnamefont{Scully}},
  \bibnamefont{and} \bibinfo{author}{\bibfnamefont{G.~R.} \bibnamefont{Welch}},
  \bibinfo{journal}{Phys. Rev. A} \textbf{\bibinfo{volume}{63}},
  \bibinfo{pages}{063802/1} (\bibinfo{year}{2001}).

\bibitem[{\citenamefont{Robinson et~al.}(1958)\citenamefont{Robinson, Ensberg,
  and Dehmelt}}]{Rob58}
\bibinfo{author}{\bibfnamefont{H.}~\bibnamefont{Robinson}},
  \bibinfo{author}{\bibfnamefont{E.}~\bibnamefont{Ensberg}}, \bibnamefont{and}
  \bibinfo{author}{\bibfnamefont{H.}~\bibnamefont{Dehmelt}},
  \bibinfo{journal}{Bull. Am. Phys. Soc.} \textbf{\bibinfo{volume}{3}},
  \bibinfo{pages}{9} (\bibinfo{year}{1958}).

\bibitem[{\citenamefont{Bouchiat and Brossel}(1966)}]{BouBro66}
\bibinfo{author}{\bibfnamefont{M.~A.} \bibnamefont{Bouchiat}} \bibnamefont{and}
  \bibinfo{author}{\bibfnamefont{J.}~\bibnamefont{Brossel}},
  \bibinfo{journal}{Phys. Rev.} \textbf{\bibinfo{volume}{147}},
  \bibinfo{pages}{41} (\bibinfo{year}{1966}).

\bibitem[{\citenamefont{Sautenkov et~al.}(2000)\citenamefont{Sautenkov, Lukin,
  Bednar, Novikova, Mikhailov, Fleischhauer, Velichansky, Welch, and
  Scully}}]{Sau2000}
\bibinfo{author}{\bibfnamefont{V.~A.} \bibnamefont{Sautenkov}},
  \bibinfo{author}{\bibfnamefont{M.~D.} \bibnamefont{Lukin}},
  \bibinfo{author}{\bibfnamefont{C.~J.} \bibnamefont{Bednar}},
  \bibinfo{author}{\bibfnamefont{I.}~\bibnamefont{Novikova}},
  \bibinfo{author}{\bibfnamefont{E.}~\bibnamefont{Mikhailov}},
  \bibinfo{author}{\bibfnamefont{M.}~\bibnamefont{Fleischhauer}},
  \bibinfo{author}{\bibfnamefont{V.~L.} \bibnamefont{Velichansky}},
  \bibinfo{author}{\bibfnamefont{G.~R.} \bibnamefont{Welch}}, \bibnamefont{and}
  \bibinfo{author}{\bibfnamefont{M.~O.} \bibnamefont{Scully}},
  \bibinfo{journal}{Phys. Rev. A} \textbf{\bibinfo{volume}{62}},
  \bibinfo{pages}{023810/1} (\bibinfo{year}{2000}).

\bibitem[{\citenamefont{Alexandrov et~al.}(2002)\citenamefont{Alexandrov,
  Balabas, Budker, English, Kimball, Li, and Yashchuk}}]{AleLIAD}
\bibinfo{author}{\bibfnamefont{E.~B.} \bibnamefont{Alexandrov}},
  \bibinfo{author}{\bibfnamefont{M.~V.} \bibnamefont{Balabas}},
  \bibinfo{author}{\bibfnamefont{D.}~\bibnamefont{Budker}},
  \bibinfo{author}{\bibfnamefont{D.}~\bibnamefont{English}},
  \bibinfo{author}{\bibfnamefont{D.~F.} \bibnamefont{Kimball}},
  \bibinfo{author}{\bibfnamefont{C.~H.} \bibnamefont{Li}}, \bibnamefont{and}
  \bibinfo{author}{\bibfnamefont{V.~V.} \bibnamefont{Yashchuk}},
  \bibinfo{journal}{Phys. Rev. A} \textbf{\bibinfo{volume}{66}},
  \bibinfo{pages}{042903/1} (\bibinfo{year}{2002}).

\bibitem[{\citenamefont{Happer and Tang}(1973)}]{Hap73}
\bibinfo{author}{\bibfnamefont{W.}~\bibnamefont{Happer}} \bibnamefont{and}
  \bibinfo{author}{\bibfnamefont{H.}~\bibnamefont{Tang}},
  \bibinfo{journal}{Phys. Rev. Lett.} \textbf{\bibinfo{volume}{31}},
  \bibinfo{pages}{273} (\bibinfo{year}{1973}).

\bibitem[{\citenamefont{Happer and Tam}(1977)}]{Hap77}
\bibinfo{author}{\bibfnamefont{W.}~\bibnamefont{Happer}} \bibnamefont{and}
  \bibinfo{author}{\bibfnamefont{A.~C.} \bibnamefont{Tam}},
  \bibinfo{journal}{Phys. Rev. A} \textbf{\bibinfo{volume}{16}},
  \bibinfo{pages}{1877} (\bibinfo{year}{1977}).

\bibitem[{\citenamefont{Dicke}(1953)}]{Dic53}
\bibinfo{author}{\bibfnamefont{R.~H.} \bibnamefont{Dicke}},
  \bibinfo{journal}{Phys. Rev.} \textbf{\bibinfo{volume}{89}},
  \bibinfo{pages}{472} (\bibinfo{year}{1953}).

\bibitem[{\citenamefont{Savukov and Romalis}(2005{\natexlab{a}})}]{Sav2005SE}
\bibinfo{author}{\bibfnamefont{I.~M.} \bibnamefont{Savukov}} \bibnamefont{and}
  \bibinfo{author}{\bibfnamefont{M.~V.} \bibnamefont{Romalis}},
  \bibinfo{journal}{Phys. Rev. A} \textbf{\bibinfo{volume}{71}},
  \bibinfo{pages}{23405} (\bibinfo{year}{2005}{\natexlab{a}}).

\bibitem[{\citenamefont{Erickson et~al.}(2000)\citenamefont{Erickson, Levron,
  Happer, Kadlecek, Chann, Anderson, and Walker}}]{Eri2000}
\bibinfo{author}{\bibfnamefont{C.~J.} \bibnamefont{Erickson}},
  \bibinfo{author}{\bibfnamefont{D.}~\bibnamefont{Levron}},
  \bibinfo{author}{\bibfnamefont{W.}~\bibnamefont{Happer}},
  \bibinfo{author}{\bibfnamefont{S.}~\bibnamefont{Kadlecek}},
  \bibinfo{author}{\bibfnamefont{B.}~\bibnamefont{Chann}},
  \bibinfo{author}{\bibfnamefont{L.~W.} \bibnamefont{Anderson}},
  \bibnamefont{and} \bibinfo{author}{\bibfnamefont{T.~G.}
  \bibnamefont{Walker}}, \bibinfo{journal}{Phys. Rev. Lett.}
  \textbf{\bibinfo{volume}{85}}, \bibinfo{pages}{4237} (\bibinfo{year}{2000}).

\bibitem[{\citenamefont{Kadlecek et~al.}(1998)\citenamefont{Kadlecek, Anderson,
  and Walker}}]{Kad98}
\bibinfo{author}{\bibfnamefont{S.}~\bibnamefont{Kadlecek}},
  \bibinfo{author}{\bibfnamefont{L.~W.} \bibnamefont{Anderson}},
  \bibnamefont{and} \bibinfo{author}{\bibfnamefont{T.~G.}
  \bibnamefont{Walker}}, \bibinfo{journal}{Phys. Rev. Lett.}
  \textbf{\bibinfo{volume}{80}}, \bibinfo{pages}{5512–5515}
  (\bibinfo{year}{1998}).

\bibitem[{\citenamefont{Allred et~al.}(2002)\citenamefont{Allred, Lyman,
  Kornack, and Romalis}}]{All2002}
\bibinfo{author}{\bibfnamefont{J.}~\bibnamefont{Allred}},
  \bibinfo{author}{\bibfnamefont{R.}~\bibnamefont{Lyman}},
  \bibinfo{author}{\bibfnamefont{T.}~\bibnamefont{Kornack}}, \bibnamefont{and}
  \bibinfo{author}{\bibfnamefont{M.}~\bibnamefont{Romalis}},
  \bibinfo{journal}{Phys. Rev. Lett.} \textbf{\bibinfo{volume}{89}},
  \bibinfo{pages}{130801} (\bibinfo{year}{2002}).

\bibitem[{\citenamefont{Seltzer and Romalis}(2004)}]{Sel2004}
\bibinfo{author}{\bibfnamefont{S.}~\bibnamefont{Seltzer}} \bibnamefont{and}
  \bibinfo{author}{\bibfnamefont{M.~V.} \bibnamefont{Romalis}},
  \bibinfo{journal}{Appl. Phys. Lett.} \textbf{\bibinfo{volume}{85}},
  \bibinfo{pages}{4804} (\bibinfo{year}{2004}).

\bibitem[{\citenamefont{Appelt et~al.}(1999)\citenamefont{Appelt,
  Ben-Amar~Baranga, Young, and Happer}}]{App99}
\bibinfo{author}{\bibfnamefont{S.}~\bibnamefont{Appelt}},
  \bibinfo{author}{\bibfnamefont{A.}~\bibnamefont{Ben-Amar~Baranga}},
  \bibinfo{author}{\bibfnamefont{A.~R.} \bibnamefont{Young}}, \bibnamefont{and}
  \bibinfo{author}{\bibfnamefont{W.}~\bibnamefont{Happer}},
  \bibinfo{journal}{Phys. Rev. A} \textbf{\bibinfo{volume}{59}},
  \bibinfo{pages}{2078} (\bibinfo{year}{1999}).

\bibitem[{\citenamefont{Jau et~al.}(2004)\citenamefont{Jau, Post, Kuzma, Braun,
  Romalis, and Happer}}]{Jau2004}
\bibinfo{author}{\bibfnamefont{Y.~Y.} \bibnamefont{Jau}},
  \bibinfo{author}{\bibfnamefont{A.~B.} \bibnamefont{Post}},
  \bibinfo{author}{\bibfnamefont{N.~N.} \bibnamefont{Kuzma}},
  \bibinfo{author}{\bibfnamefont{A.~M.} \bibnamefont{Braun}},
  \bibinfo{author}{\bibfnamefont{M.~V.} \bibnamefont{Romalis}},
  \bibnamefont{and} \bibinfo{author}{\bibfnamefont{W.}~\bibnamefont{Happer}},
  \bibinfo{journal}{Phys. Rev. Lett.} \textbf{\bibinfo{volume}{92}},
  \bibinfo{pages}{110801/1} (\bibinfo{year}{2004}).

\bibitem[{\citenamefont{Smullin et~al.}(2006)\citenamefont{Smullin, Savukov,
  Vasilakis, Ghosh, and Romalis}}]{Smu2006}
\bibinfo{author}{\bibfnamefont{S.~J.} \bibnamefont{Smullin}},
  \bibinfo{author}{\bibfnamefont{I.~M.} \bibnamefont{Savukov}},
  \bibinfo{author}{\bibfnamefont{G.}~\bibnamefont{Vasilakis}},
  \bibinfo{author}{\bibfnamefont{R.~K.} \bibnamefont{Ghosh}}, \bibnamefont{and}
  \bibinfo{author}{\bibfnamefont{M.~V.} \bibnamefont{Romalis}},
  \bibinfo{journal}{http://arxiv.org/abs/physics/0611085}
  (\bibinfo{year}{2006}).

\bibitem[{\citenamefont{Bloom}(1962)}]{Blo62}
\bibinfo{author}{\bibfnamefont{A.}~\bibnamefont{Bloom}},
  \bibinfo{journal}{Appl. Opt.} \textbf{\bibinfo{volume}{1}},
  \bibinfo{pages}{61} (\bibinfo{year}{1962}).

\bibitem[{\citenamefont{Stahler et~al.}(2001)\citenamefont{Stahler, Knappe,
  Affolderbach, Kemp, and Wynands}}]{Sta2001}
\bibinfo{author}{\bibfnamefont{M.}~\bibnamefont{Stahler}},
  \bibinfo{author}{\bibfnamefont{S.}~\bibnamefont{Knappe}},
  \bibinfo{author}{\bibfnamefont{C.}~\bibnamefont{Affolderbach}},
  \bibinfo{author}{\bibfnamefont{W.}~\bibnamefont{Kemp}}, \bibnamefont{and}
  \bibinfo{author}{\bibfnamefont{R.}~\bibnamefont{Wynands}},
  \bibinfo{journal}{Europhys. Lett.} \textbf{\bibinfo{volume}{54}},
  \bibinfo{pages}{323} (\bibinfo{year}{2001}).

\bibitem[{\citenamefont{Acosta et~al.}(2006)\citenamefont{Acosta, Ledbetter,
  Rochester, Budker, Jackson-Kimball, Hovde, Gawlik, Pustelny, and
  Zachorowski}}]{Aco2006}
\bibinfo{author}{\bibfnamefont{V.}~\bibnamefont{Acosta}},
  \bibinfo{author}{\bibfnamefont{M.~P.} \bibnamefont{Ledbetter}},
  \bibinfo{author}{\bibfnamefont{S.~M.} \bibnamefont{Rochester}},
  \bibinfo{author}{\bibfnamefont{D.}~\bibnamefont{Budker}},
  \bibinfo{author}{\bibfnamefont{D.~F.} \bibnamefont{Jackson-Kimball}},
  \bibinfo{author}{\bibfnamefont{D.~C.} \bibnamefont{Hovde}},
  \bibinfo{author}{\bibfnamefont{W.}~\bibnamefont{Gawlik}},
  \bibinfo{author}{\bibfnamefont{S.}~\bibnamefont{Pustelny}}, \bibnamefont{and}
  \bibinfo{author}{\bibfnamefont{J.}~\bibnamefont{Zachorowski}},
  \bibinfo{journal}{Phys. Rev. A} \textbf{\bibinfo{volume}{73}},
  \bibinfo{pages}{053404} (\bibinfo{year}{2006}).

\bibitem[{\citenamefont{Andreeva et~al.}(2003)\citenamefont{Andreeva,
  Bevilacqua, Biancalana, Cartaleva, Dancheva, Karaulanov, Marinelli, Mariotti,
  and Moi}}]{And2003}
\bibinfo{author}{\bibfnamefont{C.}~\bibnamefont{Andreeva}},
  \bibinfo{author}{\bibfnamefont{G.}~\bibnamefont{Bevilacqua}},
  \bibinfo{author}{\bibfnamefont{V.}~\bibnamefont{Biancalana}},
  \bibinfo{author}{\bibfnamefont{S.}~\bibnamefont{Cartaleva}},
  \bibinfo{author}{\bibfnamefont{Y.}~\bibnamefont{Dancheva}},
  \bibinfo{author}{\bibfnamefont{T.}~\bibnamefont{Karaulanov}},
  \bibinfo{author}{\bibfnamefont{C.}~\bibnamefont{Marinelli}},
  \bibinfo{author}{\bibfnamefont{E.}~\bibnamefont{Mariotti}}, \bibnamefont{and}
  \bibinfo{author}{\bibfnamefont{L.}~\bibnamefont{Moi}},
  \bibinfo{journal}{Appl. Phys. B, Lasers Opt.} \textbf{\bibinfo{volume}{76}},
  \bibinfo{pages}{667} (\bibinfo{year}{2003}).

\bibitem[{\citenamefont{Alipieva et~al.}(2005)\citenamefont{Alipieva, Andreeva,
  Avramov, Bevilaqua, Biancalana, Borisova, Breschi, Cartaleva, Dancheva,
  Gateva et~al.}}]{Ali2005}
\bibinfo{author}{\bibfnamefont{E.}~\bibnamefont{Alipieva}},
  \bibinfo{author}{\bibfnamefont{C.}~\bibnamefont{Andreeva}},
  \bibinfo{author}{\bibfnamefont{L.}~\bibnamefont{Avramov}},
  \bibinfo{author}{\bibfnamefont{G.}~\bibnamefont{Bevilaqua}},
  \bibinfo{author}{\bibfnamefont{V.}~\bibnamefont{Biancalana}},
  \bibinfo{author}{\bibfnamefont{E.}~\bibnamefont{Borisova}},
  \bibinfo{author}{\bibfnamefont{E.}~\bibnamefont{Breschi}},
  \bibinfo{author}{\bibfnamefont{S.}~\bibnamefont{Cartaleva}},
  \bibinfo{author}{\bibfnamefont{Y.}~\bibnamefont{Dancheva}},
  \bibinfo{author}{\bibfnamefont{S.}~\bibnamefont{Gateva}},
  \bibnamefont{et~al.}, \bibinfo{journal}{Proceedings of SPIE}
  \textbf{\bibinfo{volume}{5830}}, \bibinfo{pages}{170} (\bibinfo{year}{2005}).

\bibitem[{\citenamefont{Seltzer et~al.}(2006)\citenamefont{Seltzer, Meares, and
  Romalis}}]{Sel2006}
\bibinfo{author}{\bibfnamefont{S.~J.} \bibnamefont{Seltzer}},
  \bibinfo{author}{\bibfnamefont{P.~J.} \bibnamefont{Meares}},
  \bibnamefont{and} \bibinfo{author}{\bibfnamefont{M.~V.}
  \bibnamefont{Romalis}}, \bibinfo{journal}{physics/0611014}
  (\bibinfo{year}{2006}).

\bibitem[{\citenamefont{Alexandrov et~al.}(1997)\citenamefont{Alexandrov,
  Pazgalev, and Rasson}}]{Ale97}
\bibinfo{author}{\bibfnamefont{E.~B.} \bibnamefont{Alexandrov}},
  \bibinfo{author}{\bibfnamefont{A.~S.} \bibnamefont{Pazgalev}},
  \bibnamefont{and} \bibinfo{author}{\bibfnamefont{J.~L.}
  \bibnamefont{Rasson}}, \bibinfo{journal}{Opt. Spectrosk.}
  \textbf{\bibinfo{volume}{82}}, \bibinfo{pages}{14} (\bibinfo{year}{1997}).

\bibitem[{\citenamefont{Yashchuk et~al.}(2003)\citenamefont{Yashchuk, Budker,
  Gawlik, Kimball, Malakyan, and Rochester}}]{Yas2003Select}
\bibinfo{author}{\bibfnamefont{V.~V.} \bibnamefont{Yashchuk}},
  \bibinfo{author}{\bibfnamefont{D.}~\bibnamefont{Budker}},
  \bibinfo{author}{\bibfnamefont{W.}~\bibnamefont{Gawlik}},
  \bibinfo{author}{\bibfnamefont{D.~F.} \bibnamefont{Kimball}},
  \bibinfo{author}{\bibfnamefont{Y.~P.} \bibnamefont{Malakyan}},
  \bibnamefont{and} \bibinfo{author}{\bibfnamefont{S.~M.}
  \bibnamefont{Rochester}}, \bibinfo{journal}{Phys. Rev. Lett.}
  \textbf{\bibinfo{volume}{90}}, \bibinfo{pages}{253001}
  (\bibinfo{year}{2003}).

\bibitem[{\citenamefont{Pustelny
  et~al.}(2006{\natexlab{a}})\citenamefont{Pustelny, Kimball, Rochester,
  Yashchuk, Gawlik, and Budker}}]{Pus2006pp}
\bibinfo{author}{\bibfnamefont{S.}~\bibnamefont{Pustelny}},
  \bibinfo{author}{\bibfnamefont{D.~F.~J.} \bibnamefont{Kimball}},
  \bibinfo{author}{\bibfnamefont{S.~M.} \bibnamefont{Rochester}},
  \bibinfo{author}{\bibfnamefont{V.~V.} \bibnamefont{Yashchuk}},
  \bibinfo{author}{\bibfnamefont{W.}~\bibnamefont{Gawlik}}, \bibnamefont{and}
  \bibinfo{author}{\bibfnamefont{D.}~\bibnamefont{Budker}},
  \bibinfo{journal}{Phys. Rev. A} \textbf{\bibinfo{volume}{73}},
  \bibinfo{pages}{023817} (\bibinfo{year}{2006}{\natexlab{a}}).

\bibitem[{\citenamefont{Weissman}(1988)}]{Weissman88}
\bibinfo{author}{\bibfnamefont{M.~B.} \bibnamefont{Weissman}},
  \bibinfo{journal}{Rev. Mod. Phys.} \textbf{\bibinfo{volume}{60}},
  \bibinfo{pages}{537–571} (\bibinfo{year}{1988}).

\bibitem[{\citenamefont{Li et~al.}(2006)\citenamefont{Li, Wakai, and
  Walker}}]{Li2006}
\bibinfo{author}{\bibfnamefont{Z.}~\bibnamefont{Li}},
  \bibinfo{author}{\bibfnamefont{R.~T.} \bibnamefont{Wakai}}, \bibnamefont{and}
  \bibinfo{author}{\bibfnamefont{T.~G.} \bibnamefont{Walker}},
  \bibinfo{journal}{Appl. Phys. Lett.} \textbf{\bibinfo{volume}{89}},
  \bibinfo{pages}{134105} (\bibinfo{year}{2006}).

\bibitem[{\citenamefont{Alexandrov et~al.}(2004)\citenamefont{Alexandrov,
  Balabas, Kulyasov, Ivanov, Pazgalev, Rasson, Vershovskii, and
  Yakobson}}]{Ale2004VAR}
\bibinfo{author}{\bibfnamefont{E.~B.} \bibnamefont{Alexandrov}},
  \bibinfo{author}{\bibfnamefont{M.~V.} \bibnamefont{Balabas}},
  \bibinfo{author}{\bibfnamefont{V.~N.} \bibnamefont{Kulyasov}},
  \bibinfo{author}{\bibfnamefont{A.~E.} \bibnamefont{Ivanov}},
  \bibinfo{author}{\bibfnamefont{A.~S.} \bibnamefont{Pazgalev}},
  \bibinfo{author}{\bibfnamefont{J.~L.} \bibnamefont{Rasson}},
  \bibinfo{author}{\bibfnamefont{A.~K.} \bibnamefont{Vershovskii}},
  \bibnamefont{and} \bibinfo{author}{\bibfnamefont{N.~N.}
  \bibnamefont{Yakobson}}, \bibinfo{journal}{Meas. Sci. Technol. (UK)}
  \textbf{\bibinfo{volume}{15}}, \bibinfo{pages}{918} (\bibinfo{year}{2004}).

\bibitem[{\citenamefont{Gravrand et~al.}(2001)\citenamefont{Gravrand, Khokhlov,
  Mouel, and Leger}}]{Gra2001}
\bibinfo{author}{\bibfnamefont{O.}~\bibnamefont{Gravrand}},
  \bibinfo{author}{\bibfnamefont{A.}~\bibnamefont{Khokhlov}},
  \bibinfo{author}{\bibfnamefont{J.~L.~L.} \bibnamefont{Mouel}},
  \bibnamefont{and} \bibinfo{author}{\bibfnamefont{J.~M.} \bibnamefont{Leger}},
  \bibinfo{journal}{Earth Planets Space} \textbf{\bibinfo{volume}{53}},
  \bibinfo{pages}{949} (\bibinfo{year}{2001}).

\bibitem[{\citenamefont{Matsko et~al.}(2005)\citenamefont{Matsko, Strekalov,
  and Maleki}}]{Mat2005}
\bibinfo{author}{\bibfnamefont{A.~B.} \bibnamefont{Matsko}},
  \bibinfo{author}{\bibfnamefont{D.}~\bibnamefont{Strekalov}},
  \bibnamefont{and} \bibinfo{author}{\bibfnamefont{L.}~\bibnamefont{Maleki}},
  \bibinfo{journal}{Opt. Commun.} \textbf{\bibinfo{volume}{247}},
  \bibinfo{pages}{141} (\bibinfo{year}{2005}).

\bibitem[{\citenamefont{Schwindt et~al.}(2005)\citenamefont{Schwindt, Hollberg,
  and Kitching}}]{Sch2005}
\bibinfo{author}{\bibfnamefont{P.~D.~D.} \bibnamefont{Schwindt}},
  \bibinfo{author}{\bibfnamefont{L.}~\bibnamefont{Hollberg}}, \bibnamefont{and}
  \bibinfo{author}{\bibfnamefont{J.}~\bibnamefont{Kitching}},
  \bibinfo{journal}{Rev. Sci. Instrum.} \textbf{\bibinfo{volume}{76}},
  \bibinfo{pages}{126103} (\bibinfo{year}{2005}).

\bibitem[{\citenamefont{Higbie et~al.}(2006)\citenamefont{Higbie, Corsini, and
  Budker}}]{Hig2006}
\bibinfo{author}{\bibfnamefont{J.}~\bibnamefont{Higbie}},
  \bibinfo{author}{\bibfnamefont{E.}~\bibnamefont{Corsini}}, \bibnamefont{and}
  \bibinfo{author}{\bibfnamefont{D.}~\bibnamefont{Budker}},
  \bibinfo{journal}{Rev. Sci. Instrum.}  (\bibinfo{year}{2006}).

\bibitem[{\citenamefont{Bechhoefer}(2005)}]{Bec2005}
\bibinfo{author}{\bibfnamefont{J.}~\bibnamefont{Bechhoefer}},
  \bibinfo{journal}{Rev. Mod. Phys.} \textbf{\bibinfo{volume}{77}},
  \bibinfo{pages}{783} (\bibinfo{year}{2005}).

\bibitem[{\citenamefont{Rife and Boorstyn}(1974)}]{Rif74}
\bibinfo{author}{\bibfnamefont{D.~C.} \bibnamefont{Rife}} \bibnamefont{and}
  \bibinfo{author}{\bibfnamefont{R.~R.} \bibnamefont{Boorstyn}},
  \bibinfo{journal}{IEEE Trans. Inform. Theory} \textbf{\bibinfo{volume}{20}},
  \bibinfo{pages}{591} (\bibinfo{year}{1974}).

\bibitem[{\citenamefont{Balabas et~al.}(2006)\citenamefont{Balabas, Budker,
  Kitching, Schwindt, and Stalnaker}}]{Bal2006}
\bibinfo{author}{\bibfnamefont{M.~V.} \bibnamefont{Balabas}},
  \bibinfo{author}{\bibfnamefont{D.}~\bibnamefont{Budker}},
  \bibinfo{author}{\bibfnamefont{J.}~\bibnamefont{Kitching}},
  \bibinfo{author}{\bibfnamefont{P.~D.~D.} \bibnamefont{Schwindt}},
  \bibnamefont{and} \bibinfo{author}{\bibfnamefont{J.~E.}
  \bibnamefont{Stalnaker}}, \bibinfo{journal}{J. Opt. Soc. Am. B, Opt. Phys.}
  \textbf{\bibinfo{volume}{23}}, \bibinfo{pages}{1001} (\bibinfo{year}{2006}).

\bibitem[{\citenamefont{Schwindt et~al.}(2004)\citenamefont{Schwindt, Knappe,
  Shah, Hollberg, Kitching, Liew, and Moreland}}]{Sch2004}
\bibinfo{author}{\bibfnamefont{P.~D.~D.} \bibnamefont{Schwindt}},
  \bibinfo{author}{\bibfnamefont{S.}~\bibnamefont{Knappe}},
  \bibinfo{author}{\bibfnamefont{V.}~\bibnamefont{Shah}},
  \bibinfo{author}{\bibfnamefont{L.}~\bibnamefont{Hollberg}},
  \bibinfo{author}{\bibfnamefont{J.}~\bibnamefont{Kitching}},
  \bibinfo{author}{\bibfnamefont{L.~A.} \bibnamefont{Liew}}, \bibnamefont{and}
  \bibinfo{author}{\bibfnamefont{J.}~\bibnamefont{Moreland}},
  \bibinfo{journal}{Appl. Phys. Lett.} \textbf{\bibinfo{volume}{85}},
  \bibinfo{pages}{6409} (\bibinfo{year}{2004}).

\bibitem[{\citenamefont{Knappe et~al.}(2006)\citenamefont{Knappe, Schwindt,
  Gerginov, Shah, Liew, Moreland, Robinson, Hollberg, and Kitching}}]{Kna2006}
\bibinfo{author}{\bibfnamefont{S.}~\bibnamefont{Knappe}},
  \bibinfo{author}{\bibfnamefont{P.~D.~D.} \bibnamefont{Schwindt}},
  \bibinfo{author}{\bibfnamefont{V.}~\bibnamefont{Gerginov}},
  \bibinfo{author}{\bibfnamefont{V.}~\bibnamefont{Shah}},
  \bibinfo{author}{\bibfnamefont{L.}~\bibnamefont{Liew}},
  \bibinfo{author}{\bibfnamefont{J.}~\bibnamefont{Moreland}},
  \bibinfo{author}{\bibfnamefont{H.~G.} \bibnamefont{Robinson}},
  \bibinfo{author}{\bibfnamefont{L.}~\bibnamefont{Hollberg}}, \bibnamefont{and}
  \bibinfo{author}{\bibfnamefont{J.}~\bibnamefont{Kitching}},
  \bibinfo{journal}{Journal of Optics A: Pure and Applied Optics}
  \textbf{\bibinfo{volume}{8}}, \bibinfo{pages}{S318} (\bibinfo{year}{2006}).

\bibitem[{\citenamefont{Pustelny
  et~al.}(2006{\natexlab{b}})\citenamefont{Pustelny, Kimball, Rochester,
  Yashchuk, and Budker}}]{Pus2006grad}
\bibinfo{author}{\bibfnamefont{S.}~\bibnamefont{Pustelny}},
  \bibinfo{author}{\bibfnamefont{D.~F.~J.} \bibnamefont{Kimball}},
  \bibinfo{author}{\bibfnamefont{S.~M.} \bibnamefont{Rochester}},
  \bibinfo{author}{\bibfnamefont{V.~V.} \bibnamefont{Yashchuk}},
  \bibnamefont{and} \bibinfo{author}{\bibfnamefont{D.}~\bibnamefont{Budker}},
  \bibinfo{journal}{Phys. Rev. A} \textbf{\bibinfo{volume}{(physics/0608109)}}
  (\bibinfo{year}{2006}{\natexlab{b}}).

\bibitem[{\citenamefont{Fenici et~al.}(2005)\citenamefont{Fenici, Brisinda, and
  Meloni}}]{Fen2005}
\bibinfo{author}{\bibfnamefont{R.}~\bibnamefont{Fenici}},
  \bibinfo{author}{\bibfnamefont{D.}~\bibnamefont{Brisinda}}, \bibnamefont{and}
  \bibinfo{author}{\bibfnamefont{A.~M.} \bibnamefont{Meloni}},
  \bibinfo{journal}{Expert Review of Molecular Diagnostics}
  \textbf{\bibinfo{volume}{5}}, \bibinfo{pages}{291} (\bibinfo{year}{2005}).

\bibitem[{\citenamefont{H\"{a}m\"{a}l\"{a}inen
  et~al.}(1993)\citenamefont{H\"{a}m\"{a}l\"{a}inen, Hari, Ilmoniemi, Knuutila,
  and Lounasmaa}}]{Ham93}
\bibinfo{author}{\bibfnamefont{M.}~\bibnamefont{H\"{a}m\"{a}l\"{a}inen}},
  \bibinfo{author}{\bibfnamefont{R.}~\bibnamefont{Hari}},
  \bibinfo{author}{\bibfnamefont{R.~J.} \bibnamefont{Ilmoniemi}},
  \bibinfo{author}{\bibfnamefont{J.}~\bibnamefont{Knuutila}}, \bibnamefont{and}
  \bibinfo{author}{\bibfnamefont{O.~V.} \bibnamefont{Lounasmaa}},
  \bibinfo{journal}{Rev. Mod. Phys.} \textbf{\bibinfo{volume}{65}},
  \bibinfo{pages}{413–497} (\bibinfo{year}{1993}).

\bibitem[{\citenamefont{Papanicolaou et~al.}(2005)\citenamefont{Papanicolaou,
  Castillo, Billingsley-Marshall, Pataraia, and Simos}}]{Pap2005}
\bibinfo{author}{\bibfnamefont{A.~C.} \bibnamefont{Papanicolaou}},
  \bibinfo{author}{\bibfnamefont{E.~M.} \bibnamefont{Castillo}},
  \bibinfo{author}{\bibfnamefont{R.}~\bibnamefont{Billingsley-Marshall}},
  \bibinfo{author}{\bibfnamefont{E.}~\bibnamefont{Pataraia}}, \bibnamefont{and}
  \bibinfo{author}{\bibfnamefont{P.~G.} \bibnamefont{Simos}},
  \bibinfo{journal}{International Review of Neurobiology}
  \textbf{\bibinfo{volume}{68}}, \bibinfo{pages}{223} (\bibinfo{year}{2005}).

\bibitem[{\citenamefont{Livanov et~al.}(1977)\citenamefont{Livanov, Kozlov,
  Korinevskii, Markin, Sinelnikova, and Kholodov}}]{Liv77}
\bibinfo{author}{\bibfnamefont{M.~N.} \bibnamefont{Livanov}},
  \bibinfo{author}{\bibfnamefont{A.~N.} \bibnamefont{Kozlov}},
  \bibinfo{author}{\bibfnamefont{A.~V.} \bibnamefont{Korinevskii}},
  \bibinfo{author}{\bibfnamefont{V.~P.} \bibnamefont{Markin}},
  \bibinfo{author}{\bibfnamefont{S.~E.} \bibnamefont{Sinelnikova}},
  \bibnamefont{and} \bibinfo{author}{\bibfnamefont{I.~A.}
  \bibnamefont{Kholodov}}, \bibinfo{journal}{Doklady Akademii Nauk SSSR}
  \textbf{\bibinfo{volume}{238}}, \bibinfo{pages}{253} (\bibinfo{year}{1977}).

\bibitem[{\citenamefont{Bison et~al.}(2003)\citenamefont{Bison, Wynands, and
  Weis}}]{Bis2003}
\bibinfo{author}{\bibfnamefont{G.}~\bibnamefont{Bison}},
  \bibinfo{author}{\bibfnamefont{R.}~\bibnamefont{Wynands}}, \bibnamefont{and}
  \bibinfo{author}{\bibfnamefont{A.}~\bibnamefont{Weis}},
  \bibinfo{journal}{Appl. Phys. B.} \textbf{\bibinfo{volume}{76}},
  \bibinfo{pages}{325} (\bibinfo{year}{2003}).

\bibitem[{\citenamefont{Xia et~al.}(2006)\citenamefont{Xia, Baranga, Hoffman,
  and Romalis}}]{Xia2006}
\bibinfo{author}{\bibfnamefont{H.}~\bibnamefont{Xia}},
  \bibinfo{author}{\bibfnamefont{A.~B.} \bibnamefont{Baranga}},
  \bibinfo{author}{\bibfnamefont{D.}~\bibnamefont{Hoffman}}, \bibnamefont{and}
  \bibinfo{author}{\bibfnamefont{M.~V.} \bibnamefont{Romalis}},
  \bibinfo{journal}{Appl. Phys. Lett.}  (\bibinfo{year}{2006}).

\bibitem[{\citenamefont{Murthy et~al.}(1989)\citenamefont{Murthy, Krause, Li,
  and Hunter}}]{Mur89}
\bibinfo{author}{\bibfnamefont{S.~A.} \bibnamefont{Murthy}},
  \bibinfo{author}{\bibfnamefont{J.}~\bibnamefont{Krause}, \bibfnamefont{D.}},
  \bibinfo{author}{\bibfnamefont{Z.~L.} \bibnamefont{Li}}, \bibnamefont{and}
  \bibinfo{author}{\bibfnamefont{L.~R.} \bibnamefont{Hunter}},
  \bibinfo{journal}{Phys. Rev. Lett.} \textbf{\bibinfo{volume}{63}},
  \bibinfo{pages}{965} (\bibinfo{year}{1989}).

\bibitem[{\citenamefont{Berglund et~al.}(1995)\citenamefont{Berglund, Hunter,
  Krause, Prigge, Ronfeldt, and Lamoreaux}}]{Ber95}
\bibinfo{author}{\bibfnamefont{C.~J.} \bibnamefont{Berglund}},
  \bibinfo{author}{\bibfnamefont{L.~R.} \bibnamefont{Hunter}},
  \bibinfo{author}{\bibfnamefont{J.}~\bibnamefont{Krause}, \bibfnamefont{D.}},
  \bibinfo{author}{\bibfnamefont{E.~O.} \bibnamefont{Prigge}},
  \bibinfo{author}{\bibfnamefont{M.~S.} \bibnamefont{Ronfeldt}},
  \bibnamefont{and} \bibinfo{author}{\bibfnamefont{S.~K.}
  \bibnamefont{Lamoreaux}}, \bibinfo{journal}{Phys. Rev. Lett.}
  \textbf{\bibinfo{volume}{75}}, \bibinfo{pages}{1879} (\bibinfo{year}{1995}).

\bibitem[{\citenamefont{Youdin et~al.}(1996)\citenamefont{Youdin, Krause,
  Jagannathan, Hunter, and Lamoreaux}}]{You96}
\bibinfo{author}{\bibfnamefont{A.~N.} \bibnamefont{Youdin}},
  \bibinfo{author}{\bibfnamefont{J.}~\bibnamefont{Krause}, \bibfnamefont{D.}},
  \bibinfo{author}{\bibfnamefont{K.}~\bibnamefont{Jagannathan}},
  \bibinfo{author}{\bibfnamefont{L.~R.} \bibnamefont{Hunter}},
  \bibnamefont{and} \bibinfo{author}{\bibfnamefont{S.~K.}
  \bibnamefont{Lamoreaux}}, \bibinfo{journal}{Phys. Rev. Lett.}
  \textbf{\bibinfo{volume}{77}}, \bibinfo{pages}{2170} (\bibinfo{year}{1996}).

\bibitem[{\citenamefont{Gilles et~al.}(2003)\citenamefont{Gilles, Monfort, and
  Hamel}}]{Gil2003}
\bibinfo{author}{\bibfnamefont{H.}~\bibnamefont{Gilles}},
  \bibinfo{author}{\bibfnamefont{Y.}~\bibnamefont{Monfort}}, \bibnamefont{and}
  \bibinfo{author}{\bibfnamefont{J.}~\bibnamefont{Hamel}},
  \bibinfo{journal}{Rev. Sci. Instrum.} \textbf{\bibinfo{volume}{74}},
  \bibinfo{pages}{4515} (\bibinfo{year}{2003}).

\bibitem[{\citenamefont{Romalis et~al.}(2001)\citenamefont{Romalis, Griffith,
  Jacobs, and Fortson}}]{Rom2001PRL}
\bibinfo{author}{\bibfnamefont{M.~V.} \bibnamefont{Romalis}},
  \bibinfo{author}{\bibfnamefont{W.~C.} \bibnamefont{Griffith}},
  \bibinfo{author}{\bibfnamefont{J.~P.} \bibnamefont{Jacobs}},
  \bibnamefont{and} \bibinfo{author}{\bibfnamefont{E.~N.}
  \bibnamefont{Fortson}}, \bibinfo{journal}{Phys. Rev. Lett.}
  \textbf{\bibinfo{volume}{86}}, \bibinfo{pages}{2505} (\bibinfo{year}{2001}).

\bibitem[{\citenamefont{Bear et~al.}(2000)\citenamefont{Bear, Stoner,
  Walsworth, Kostelecký, and Lane}}]{Bea2000}
\bibinfo{author}{\bibfnamefont{D.}~\bibnamefont{Bear}},
  \bibinfo{author}{\bibfnamefont{R.~E.} \bibnamefont{Stoner}},
  \bibinfo{author}{\bibfnamefont{R.~L.} \bibnamefont{Walsworth}},
  \bibinfo{author}{\bibfnamefont{V.~A.} \bibnamefont{Kostelecký}},
  \bibnamefont{and} \bibinfo{author}{\bibfnamefont{C.~D.} \bibnamefont{Lane}},
  \bibinfo{journal}{Phys. Rev. Lett.} \textbf{\bibinfo{volume}{85}},
  \bibinfo{pages}{5038} (\bibinfo{year}{2000}).

\bibitem[{\citenamefont{Bear et~al.}(2002)\citenamefont{Bear, Stoner,
  Walsworth, Kostelecký, and Lane}}]{Bea2002}
\bibinfo{author}{\bibfnamefont{D.}~\bibnamefont{Bear}},
  \bibinfo{author}{\bibfnamefont{R.~E.} \bibnamefont{Stoner}},
  \bibinfo{author}{\bibfnamefont{R.~L.} \bibnamefont{Walsworth}},
  \bibinfo{author}{\bibfnamefont{V.~A.} \bibnamefont{Kostelecký}},
  \bibnamefont{and} \bibinfo{author}{\bibfnamefont{C.~D.} \bibnamefont{Lane}},
  \bibinfo{journal}{Phys. Rev. Lett.} \textbf{\bibinfo{volume}{89}},
  \bibinfo{pages}{209902} (\bibinfo{year}{2002}).

\bibitem[{\citenamefont{Chin et~al.}(2001)\citenamefont{Chin, Leiber, Vuletic,
  Kerman, and Chu}}]{Chi2001}
\bibinfo{author}{\bibfnamefont{C.}~\bibnamefont{Chin}},
  \bibinfo{author}{\bibfnamefont{V.}~\bibnamefont{Leiber}},
  \bibinfo{author}{\bibfnamefont{V.}~\bibnamefont{Vuletic}},
  \bibinfo{author}{\bibfnamefont{A.~J.} \bibnamefont{Kerman}},
  \bibnamefont{and} \bibinfo{author}{\bibfnamefont{S.}~\bibnamefont{Chu}},
  \bibinfo{journal}{Phys. Rev. A} \textbf{\bibinfo{volume}{63}},
  \bibinfo{pages}{0334011} (\bibinfo{year}{2001}).

\bibitem[{\citenamefont{Amini et~al.}(2006)\citenamefont{Amini, Jr., and
  Gould}}]{Ami2006}
\bibinfo{author}{\bibfnamefont{J.~M.} \bibnamefont{Amini}},
  \bibinfo{author}{\bibfnamefont{C.~T.~M.} \bibnamefont{Jr.}},
  \bibnamefont{and} \bibinfo{author}{\bibfnamefont{H.}~\bibnamefont{Gould}},
  \bibinfo{journal}{http://arxiv.org/physics/0602011}  (\bibinfo{year}{2006}).

\bibitem[{\citenamefont{Lamoreaux}(2002)}]{Lam2002}
\bibinfo{author}{\bibfnamefont{S.~K.} \bibnamefont{Lamoreaux}},
  \bibinfo{journal}{Phys. Rev. A} \textbf{\bibinfo{volume}{66}},
  \bibinfo{pages}{022109} (\bibinfo{year}{2002}).

\bibitem[{\citenamefont{Budker et~al.}(2006)\citenamefont{Budker, Lamoreaux,
  Sushkov, and Sushkov}}]{Bud2006CM}
\bibinfo{author}{\bibfnamefont{D.}~\bibnamefont{Budker}},
  \bibinfo{author}{\bibfnamefont{S.~K.} \bibnamefont{Lamoreaux}},
  \bibinfo{author}{\bibfnamefont{A.~O.} \bibnamefont{Sushkov}},
  \bibnamefont{and} \bibinfo{author}{\bibfnamefont{O.~P.}
  \bibnamefont{Sushkov}}, \bibinfo{journal}{Phys. Rev. A}
  \textbf{\bibinfo{volume}{73}}, \bibinfo{pages}{022107}
  (\bibinfo{year}{2006}).

\bibitem[{\citenamefont{Ness}(1970)}]{Nes70}
\bibinfo{author}{\bibfnamefont{N.~F.} \bibnamefont{Ness}},
  \bibinfo{journal}{Space Sci. Rev. (Netherlands)}
  \textbf{\bibinfo{volume}{11}}, \bibinfo{pages}{459} (\bibinfo{year}{1970}).

\bibitem[{\citenamefont{Acuna}(1997)}]{Acu97}
\bibinfo{author}{\bibfnamefont{M.~H.} \bibnamefont{Acuna}}, in
  \emph{\bibinfo{booktitle}{Encyclopedia of planetary sciences}}, edited by
  \bibinfo{editor}{\bibfnamefont{J.~H.} \bibnamefont{Shirley}}
  \bibnamefont{and} \bibinfo{editor}{\bibfnamefont{R.~W.}
  \bibnamefont{Fairbridge}} (\bibinfo{publisher}{Chapman \& Hall},
  \bibinfo{address}{London}, \bibinfo{year}{1997}).

\bibitem[{\citenamefont{Balogh}(1988)}]{Bal88}
\bibinfo{author}{\bibfnamefont{A.}~\bibnamefont{Balogh}}, in
  \emph{\bibinfo{booktitle}{IEE Colloquium on `Satellite Instrumentation'}}
  (\bibinfo{publisher}{IEE}, \bibinfo{address}{London, UK},
  \bibinfo{year}{1988}), vol. \bibinfo{volume}{Digest No.12}, pp.
  \bibinfo{pages}{2/1--3}.

\bibitem[{\citenamefont{Southwood et~al.}(1992)\citenamefont{Southwood, Balogh,
  and Smith}}]{Sou92}
\bibinfo{author}{\bibfnamefont{D.~J.} \bibnamefont{Southwood}},
  \bibinfo{author}{\bibfnamefont{A.}~\bibnamefont{Balogh}}, \bibnamefont{and}
  \bibinfo{author}{\bibfnamefont{E.~J.} \bibnamefont{Smith}},
  \bibinfo{journal}{J. Br. Interplanet. Soc. (UK)}
  \textbf{\bibinfo{volume}{45}}, \bibinfo{pages}{371} (\bibinfo{year}{1992}).

\bibitem[{\citenamefont{Dunlop et~al.}(1999)\citenamefont{Dunlop, Dougherty,
  Kellock, and Southwood}}]{Dun99}
\bibinfo{author}{\bibfnamefont{M.~W.} \bibnamefont{Dunlop}},
  \bibinfo{author}{\bibfnamefont{M.~K.} \bibnamefont{Dougherty}},
  \bibinfo{author}{\bibfnamefont{S.}~\bibnamefont{Kellock}}, \bibnamefont{and}
  \bibinfo{author}{\bibfnamefont{D.~J.} \bibnamefont{Southwood}},
  \bibinfo{journal}{Planet. Space Sci. (UK)} \textbf{\bibinfo{volume}{47}},
  \bibinfo{pages}{1389} (\bibinfo{year}{1999}).

\bibitem[{\citenamefont{Dougherty et~al.}(2005)\citenamefont{Dougherty,
  Achilleos, Andre, Arridge, Balogh, Bertucci, Burton, Cowley, Erdos, Giampieri
  et~al.}}]{Dou2005}
\bibinfo{author}{\bibfnamefont{M.~K.} \bibnamefont{Dougherty}},
  \bibinfo{author}{\bibfnamefont{N.}~\bibnamefont{Achilleos}},
  \bibinfo{author}{\bibfnamefont{N.}~\bibnamefont{Andre}},
  \bibinfo{author}{\bibfnamefont{C.~S.} \bibnamefont{Arridge}},
  \bibinfo{author}{\bibfnamefont{A.}~\bibnamefont{Balogh}},
  \bibinfo{author}{\bibfnamefont{C.}~\bibnamefont{Bertucci}},
  \bibinfo{author}{\bibfnamefont{M.~E.} \bibnamefont{Burton}},
  \bibinfo{author}{\bibfnamefont{S.~W.~H.} \bibnamefont{Cowley}},
  \bibinfo{author}{\bibfnamefont{G.}~\bibnamefont{Erdos}},
  \bibinfo{author}{\bibfnamefont{G.}~\bibnamefont{Giampieri}},
  \bibnamefont{et~al.}, \bibinfo{journal}{Science}
  \textbf{\bibinfo{volume}{307}}, \bibinfo{pages}{1266} (\bibinfo{year}{2005}).

\bibitem[{\citenamefont{Dougherty et~al.}(2006)\citenamefont{Dougherty,
  Khurana, Neubauer, Russell, Saur, Leisner, and Burton}}]{Dou2006}
\bibinfo{author}{\bibfnamefont{M.~K.} \bibnamefont{Dougherty}},
  \bibinfo{author}{\bibfnamefont{K.~K.} \bibnamefont{Khurana}},
  \bibinfo{author}{\bibfnamefont{F.~M.} \bibnamefont{Neubauer}},
  \bibinfo{author}{\bibfnamefont{C.~T.} \bibnamefont{Russell}},
  \bibinfo{author}{\bibfnamefont{J.}~\bibnamefont{Saur}},
  \bibinfo{author}{\bibfnamefont{J.~S.} \bibnamefont{Leisner}},
  \bibnamefont{and} \bibinfo{author}{\bibfnamefont{M.~E.}
  \bibnamefont{Burton}}, \bibinfo{journal}{Science}
  \textbf{\bibinfo{volume}{311}}, \bibinfo{pages}{1406} (\bibinfo{year}{2006}).

\bibitem[{\citenamefont{Slocum et~al.}(2002)\citenamefont{Slocum, Kuhlman,
  Ryan, and King}}]{Slo2002}
\bibinfo{author}{\bibfnamefont{R.~E.} \bibnamefont{Slocum}},
  \bibinfo{author}{\bibfnamefont{G.}~\bibnamefont{Kuhlman}},
  \bibinfo{author}{\bibfnamefont{L.}~\bibnamefont{Ryan}}, \bibnamefont{and}
  \bibinfo{author}{\bibfnamefont{D.}~\bibnamefont{King}}, in
  \emph{\bibinfo{booktitle}{Oceans 2002}} (\bibinfo{publisher}{IEEE. Conference
  Proceedings (Cat. No.02CH37362).}, \bibinfo{year}{2002}),
  vol.~\bibinfo{volume}{2}, pp. \bibinfo{pages}{945--51}.

\bibitem[{\citenamefont{McGregor}(1987)}]{McG87}
\bibinfo{author}{\bibfnamefont{D.~D.} \bibnamefont{McGregor}},
  \bibinfo{journal}{Rev. Sci. Instrum.} \textbf{\bibinfo{volume}{58}},
  \bibinfo{pages}{1067} (\bibinfo{year}{1987}).

\bibitem[{\citenamefont{Burlaga et~al.}(2005)\citenamefont{Burlaga, Ness, Wang,
  Richardson, McDonald, and Stone}}]{Bur2005}
\bibinfo{author}{\bibfnamefont{L.~F.} \bibnamefont{Burlaga}},
  \bibinfo{author}{\bibfnamefont{N.~F.} \bibnamefont{Ness}},
  \bibinfo{author}{\bibfnamefont{C.}~\bibnamefont{Wang}},
  \bibinfo{author}{\bibfnamefont{J.~D.} \bibnamefont{Richardson}},
  \bibinfo{author}{\bibfnamefont{F.~B.} \bibnamefont{McDonald}},
  \bibnamefont{and} \bibinfo{author}{\bibfnamefont{E.~C.} \bibnamefont{Stone}},
  \bibinfo{journal}{Astrophysical Journal} \textbf{\bibinfo{volume}{618}},
  \bibinfo{pages}{1074–1078} (\bibinfo{year}{2005}).

\bibitem[{\citenamefont{Greenberg}(1998)}]{Gre98}
\bibinfo{author}{\bibfnamefont{Y.~S.} \bibnamefont{Greenberg}},
  \bibinfo{journal}{Rev. Mod. Phys.} \textbf{\bibinfo{volume}{70}},
  \bibinfo{pages}{175} (\bibinfo{year}{1998}).

\bibitem[{\citenamefont{Cohen-Tannoudji
  et~al.}(1969)\citenamefont{Cohen-Tannoudji, DuPont-Roc, Haroche, and
  Laloe}}]{Coh69b}
\bibinfo{author}{\bibfnamefont{C.}~\bibnamefont{Cohen-Tannoudji}},
  \bibinfo{author}{\bibfnamefont{J.}~\bibnamefont{DuPont-Roc}},
  \bibinfo{author}{\bibfnamefont{S.}~\bibnamefont{Haroche}}, \bibnamefont{and}
  \bibinfo{author}{\bibfnamefont{F.}~\bibnamefont{Laloe}},
  \bibinfo{journal}{Phys. Rev. Lett.} \textbf{\bibinfo{volume}{22}},
  \bibinfo{pages}{758} (\bibinfo{year}{1969}).

\bibitem[{\citenamefont{Yashchuk et~al.}(2004)\citenamefont{Yashchuk, Granwehr,
  Kimball, Rochester, Trabesinger, Urban, Budker, and Pines}}]{Yas2004}
\bibinfo{author}{\bibfnamefont{V.~V.} \bibnamefont{Yashchuk}},
  \bibinfo{author}{\bibfnamefont{J.}~\bibnamefont{Granwehr}},
  \bibinfo{author}{\bibfnamefont{D.~F.} \bibnamefont{Kimball}},
  \bibinfo{author}{\bibfnamefont{S.~M.} \bibnamefont{Rochester}},
  \bibinfo{author}{\bibfnamefont{A.~H.} \bibnamefont{Trabesinger}},
  \bibinfo{author}{\bibfnamefont{J.~T.} \bibnamefont{Urban}},
  \bibinfo{author}{\bibfnamefont{D.}~\bibnamefont{Budker}}, \bibnamefont{and}
  \bibinfo{author}{\bibfnamefont{A.}~\bibnamefont{Pines}},
  \bibinfo{journal}{Phys. Rev. Lett.} \textbf{\bibinfo{volume}{93}},
  \bibinfo{pages}{160801} (\bibinfo{year}{2004}).

\bibitem[{\citenamefont{Savukov and Romalis}(2005{\natexlab{b}})}]{Sav2005NMR}
\bibinfo{author}{\bibfnamefont{I.~M.} \bibnamefont{Savukov}} \bibnamefont{and}
  \bibinfo{author}{\bibfnamefont{M.~V.} \bibnamefont{Romalis}},
  \bibinfo{journal}{Phys. Rev. Lett.} \textbf{\bibinfo{volume}{94}},
  \bibinfo{pages}{1230011} (\bibinfo{year}{2005}{\natexlab{b}}).

\bibitem[{\citenamefont{Moul\'{e} et~al.}(2003)\citenamefont{Moul\'{e}, Spence,
  Han, Seeley, Pierce, Saxena, and Pines}}]{Mou2003}
\bibinfo{author}{\bibfnamefont{A.~J.} \bibnamefont{Moul\'{e}}},
  \bibinfo{author}{\bibfnamefont{M.~M.} \bibnamefont{Spence}},
  \bibinfo{author}{\bibfnamefont{S.}~\bibnamefont{Han}},
  \bibinfo{author}{\bibfnamefont{J.~A.} \bibnamefont{Seeley}},
  \bibinfo{author}{\bibfnamefont{K.~L.} \bibnamefont{Pierce}},
  \bibinfo{author}{\bibfnamefont{S.}~\bibnamefont{Saxena}}, \bibnamefont{and}
  \bibinfo{author}{\bibfnamefont{A.}~\bibnamefont{Pines}},
  \bibinfo{journal}{Proc. Natl. Acad. Sci. U. S. A}
  \textbf{\bibinfo{volume}{100}}, \bibinfo{pages}{9122} (\bibinfo{year}{2003}).

\bibitem[{\citenamefont{Xu et~al.}(2006{\natexlab{a}})\citenamefont{Xu,
  Rochester, Yashchuk, Donaldson, and Budker}}]{Xu2006RSI}
\bibinfo{author}{\bibfnamefont{S.}~\bibnamefont{Xu}},
  \bibinfo{author}{\bibfnamefont{S.~M.} \bibnamefont{Rochester}},
  \bibinfo{author}{\bibfnamefont{V.~V.} \bibnamefont{Yashchuk}},
  \bibinfo{author}{\bibfnamefont{M.~H.} \bibnamefont{Donaldson}},
  \bibnamefont{and} \bibinfo{author}{\bibfnamefont{D.}~\bibnamefont{Budker}},
  \bibinfo{journal}{Rev. Sci. Instrum.} \textbf{\bibinfo{volume}{77}},
  \bibinfo{pages}{083106} (\bibinfo{year}{2006}{\natexlab{a}}).

\bibitem[{\citenamefont{Xu et~al.}(2006{\natexlab{b}})\citenamefont{Xu,
  Yashchuk, Donaldson, Rochester, Budker, and Pines}}]{Xu2006IMAG}
\bibinfo{author}{\bibfnamefont{S.}~\bibnamefont{Xu}},
  \bibinfo{author}{\bibfnamefont{V.~V.} \bibnamefont{Yashchuk}},
  \bibinfo{author}{\bibfnamefont{M.~H.} \bibnamefont{Donaldson}},
  \bibinfo{author}{\bibfnamefont{S.~M.} \bibnamefont{Rochester}},
  \bibinfo{author}{\bibfnamefont{D.}~\bibnamefont{Budker}}, \bibnamefont{and}
  \bibinfo{author}{\bibfnamefont{A.}~\bibnamefont{Pines}},
  \bibinfo{journal}{Proc. Natl. Acad. Sci. U. S. A} p.
  \bibinfo{pages}{0605396103} (\bibinfo{year}{2006}{\natexlab{b}}).

\bibitem[{\citenamefont{Schaefer et~al.}(1989)\citenamefont{Schaefer, Cates,
  Chien, Gonatas, Happer, and Walker}}]{Sch89PRA}
\bibinfo{author}{\bibfnamefont{S.~R.} \bibnamefont{Schaefer}},
  \bibinfo{author}{\bibfnamefont{G.~D.} \bibnamefont{Cates}},
  \bibinfo{author}{\bibfnamefont{T.-R.} \bibnamefont{Chien}},
  \bibinfo{author}{\bibfnamefont{D.}~\bibnamefont{Gonatas}},
  \bibinfo{author}{\bibfnamefont{W.}~\bibnamefont{Happer}}, \bibnamefont{and}
  \bibinfo{author}{\bibfnamefont{T.~G.} \bibnamefont{Walker}},
  \bibinfo{journal}{Phys. Rev. A} \textbf{\bibinfo{volume}{39}},
  \bibinfo{pages}{5613} (\bibinfo{year}{1989}).

\bibitem[{\citenamefont{Woodman et~al.}(1987)\citenamefont{Woodman, Franks, and
  Richards}}]{Woo87}
\bibinfo{author}{\bibfnamefont{K.~F.} \bibnamefont{Woodman}},
  \bibinfo{author}{\bibfnamefont{P.~W.} \bibnamefont{Franks}},
  \bibnamefont{and} \bibinfo{author}{\bibfnamefont{M.~D.}
  \bibnamefont{Richards}}, \bibinfo{journal}{Journal of Navigation}
  \textbf{\bibinfo{volume}{40}}, \bibinfo{pages}{366} (\bibinfo{year}{1987}).

\bibitem[{\citenamefont{Kornack et~al.}(2005)\citenamefont{Kornack, Ghosh, and
  Romalis}}]{Kor2005}
\bibinfo{author}{\bibfnamefont{T.~W.} \bibnamefont{Kornack}},
  \bibinfo{author}{\bibfnamefont{R.~K.} \bibnamefont{Ghosh}}, \bibnamefont{and}
  \bibinfo{author}{\bibfnamefont{M.~V.} \bibnamefont{Romalis}},
  \bibinfo{journal}{Phys. Rev. Lett.} \textbf{\bibinfo{volume}{95}},
  \bibinfo{pages}{230801} (\bibinfo{year}{2005}).

\bibitem[{\citenamefont{Budker et~al.}(2003)\citenamefont{Budker, Kimball,
  Rochester, and Urban}}]{Bud2003NQR}
\bibinfo{author}{\bibfnamefont{D.}~\bibnamefont{Budker}},
  \bibinfo{author}{\bibfnamefont{D.~F.} \bibnamefont{Kimball}},
  \bibinfo{author}{\bibfnamefont{S.~M.} \bibnamefont{Rochester}},
  \bibnamefont{and} \bibinfo{author}{\bibfnamefont{J.~T.} \bibnamefont{Urban}},
  \bibinfo{journal}{Chem. Phys. Lett.} \textbf{\bibinfo{volume}{378}},
  \bibinfo{pages}{440} (\bibinfo{year}{2003}).

\bibitem[{\citenamefont{Garroway et~al.}(2001)\citenamefont{Garroway, Buess,
  Miller, Suits, Hibbs, Barrall, Matthews, and Burnett}}]{Gar2001}
\bibinfo{author}{\bibfnamefont{A.~N.} \bibnamefont{Garroway}},
  \bibinfo{author}{\bibfnamefont{M.~L.} \bibnamefont{Buess}},
  \bibinfo{author}{\bibfnamefont{J.~B.} \bibnamefont{Miller}},
  \bibinfo{author}{\bibfnamefont{B.~H.} \bibnamefont{Suits}},
  \bibinfo{author}{\bibfnamefont{A.~D.} \bibnamefont{Hibbs}},
  \bibinfo{author}{\bibfnamefont{G.~A.} \bibnamefont{Barrall}},
  \bibinfo{author}{\bibfnamefont{R.}~\bibnamefont{Matthews}}, \bibnamefont{and}
  \bibinfo{author}{\bibfnamefont{L.~J.} \bibnamefont{Burnett}},
  \bibinfo{journal}{IEEE Trans. Geosci. Remote Sens. (USA)}
  \textbf{\bibinfo{volume}{39}}, \bibinfo{pages}{1108} (\bibinfo{year}{2001}).

\bibitem[{\citenamefont{Ledbetter et~al.}(2006)\citenamefont{Ledbetter, Acosta,
  Rochester, Budker, Pustelny, and Yashchuk}}]{Led2006}
\bibinfo{author}{\bibfnamefont{M.~P.} \bibnamefont{Ledbetter}},
  \bibinfo{author}{\bibfnamefont{V.~M.} \bibnamefont{Acosta}},
  \bibinfo{author}{\bibfnamefont{S.~M.} \bibnamefont{Rochester}},
  \bibinfo{author}{\bibfnamefont{D.}~\bibnamefont{Budker}},
  \bibinfo{author}{\bibfnamefont{S.}~\bibnamefont{Pustelny}}, \bibnamefont{and}
  \bibinfo{author}{\bibfnamefont{V.~V.} \bibnamefont{Yashchuk}},
  \bibinfo{journal}{physics/0609196}  (\bibinfo{year}{2006}).

\bibitem[{\citenamefont{Lee et~al.}(2006)\citenamefont{Lee, Sauer, Seltzer,
  Alem, and Romalis}}]{Lee2006}
\bibinfo{author}{\bibfnamefont{S.-K.} \bibnamefont{Lee}},
  \bibinfo{author}{\bibfnamefont{K.~L.} \bibnamefont{Sauer}},
  \bibinfo{author}{\bibfnamefont{S.~J.} \bibnamefont{Seltzer}},
  \bibinfo{author}{\bibfnamefont{O.}~\bibnamefont{Alem}}, \bibnamefont{and}
  \bibinfo{author}{\bibfnamefont{M.~V.} \bibnamefont{Romalis}},
  \bibinfo{journal}{Appl. Phys. Lett.} \textbf{\bibinfo{volume}{(in press)}}
  (\bibinfo{year}{2006}).

\bibitem[{\citenamefont{Savukov et~al.}(2006)\citenamefont{Savukov, Seltzer,
  and Romalis}}]{Sav2006NMR_rf}
\bibinfo{author}{\bibfnamefont{I.~M.} \bibnamefont{Savukov}},
  \bibinfo{author}{\bibfnamefont{S.~J.} \bibnamefont{Seltzer}},
  \bibnamefont{and} \bibinfo{author}{\bibfnamefont{M.~V.}
  \bibnamefont{Romalis}}, \bibinfo{journal}{Submitted}  (\bibinfo{year}{2006}).

\bibitem[{\citenamefont{O'Hara et~al.}(1999)\citenamefont{O'Hara, Granade,
  Gehm, Savard, Bali, Freed, and Thomas}}]{OHa99}
\bibinfo{author}{\bibfnamefont{K.~M.} \bibnamefont{O'Hara}},
  \bibinfo{author}{\bibfnamefont{S.~R.} \bibnamefont{Granade}},
  \bibinfo{author}{\bibfnamefont{M.~E.} \bibnamefont{Gehm}},
  \bibinfo{author}{\bibfnamefont{T.~A.} \bibnamefont{Savard}},
  \bibinfo{author}{\bibfnamefont{S.}~\bibnamefont{Bali}},
  \bibinfo{author}{\bibfnamefont{C.}~\bibnamefont{Freed}}, \bibnamefont{and}
  \bibinfo{author}{\bibfnamefont{J.~E.} \bibnamefont{Thomas}},
  \bibinfo{journal}{Phys. Rev. Lett.} \textbf{\bibinfo{volume}{82}},
  \bibinfo{pages}{4204} (\bibinfo{year}{1999}).

\bibitem[{\citenamefont{Freeman and Choi}(2001)}]{Fre2001}
\bibinfo{author}{\bibfnamefont{M.~R.} \bibnamefont{Freeman}} \bibnamefont{and}
  \bibinfo{author}{\bibfnamefont{B.~C.} \bibnamefont{Choi}},
  \bibinfo{journal}{Science} \textbf{\bibinfo{volume}{294}},
  \bibinfo{pages}{1484} (\bibinfo{year}{2001}).

\bibitem[{\citenamefont{Chatraphorn et~al.}(2000)\citenamefont{Chatraphorn,
  Fleet, Wellstood, Knauss, and Eiles}}]{Cha2000}
\bibinfo{author}{\bibfnamefont{S.}~\bibnamefont{Chatraphorn}},
  \bibinfo{author}{\bibfnamefont{E.~F.} \bibnamefont{Fleet}},
  \bibinfo{author}{\bibfnamefont{F.~C.} \bibnamefont{Wellstood}},
  \bibinfo{author}{\bibfnamefont{L.~A.} \bibnamefont{Knauss}},
  \bibnamefont{and} \bibinfo{author}{\bibfnamefont{T.~M.} \bibnamefont{Eiles}},
  \bibinfo{journal}{Appl. Phys. Lett.} \textbf{\bibinfo{volume}{76}},
  \bibinfo{pages}{2304} (\bibinfo{year}{2000}).

\bibitem[{\citenamefont{Wildermuth et~al.}(2005)\citenamefont{Wildermuth,
  Hofferberth, Lesanovsky, Haller, Andersson, Groth, Bar-Joseph, Kruger, and
  Scmiedmayer}}]{Wil2005}
\bibinfo{author}{\bibfnamefont{S.}~\bibnamefont{Wildermuth}},
  \bibinfo{author}{\bibfnamefont{S.}~\bibnamefont{Hofferberth}},
  \bibinfo{author}{\bibfnamefont{I.}~\bibnamefont{Lesanovsky}},
  \bibinfo{author}{\bibfnamefont{E.}~\bibnamefont{Haller}},
  \bibinfo{author}{\bibfnamefont{L.~M.} \bibnamefont{Andersson}},
  \bibinfo{author}{\bibfnamefont{S.}~\bibnamefont{Groth}},
  \bibinfo{author}{\bibfnamefont{I.}~\bibnamefont{Bar-Joseph}},
  \bibinfo{author}{\bibfnamefont{P.}~\bibnamefont{Kruger}}, \bibnamefont{and}
  \bibinfo{author}{\bibfnamefont{J.}~\bibnamefont{Scmiedmayer}},
  \bibinfo{journal}{Nature} \textbf{\bibinfo{volume}{435}},
  \bibinfo{pages}{440} (\bibinfo{year}{2005}).

\bibitem[{\citenamefont{Wildermuth et~al.}(2006)\citenamefont{Wildermuth,
  Hofferberth, Lesanovsky, Groth, Krüger, Schmiedmayer, and
  Bar-Joseph}}]{Wil2006}
\bibinfo{author}{\bibfnamefont{S.}~\bibnamefont{Wildermuth}},
  \bibinfo{author}{\bibfnamefont{S.}~\bibnamefont{Hofferberth}},
  \bibinfo{author}{\bibfnamefont{I.}~\bibnamefont{Lesanovsky}},
  \bibinfo{author}{\bibfnamefont{S.}~\bibnamefont{Groth}},
  \bibinfo{author}{\bibfnamefont{P.}~\bibnamefont{Krüger}},
  \bibinfo{author}{\bibfnamefont{J.}~\bibnamefont{Schmiedmayer}},
  \bibnamefont{and}
  \bibinfo{author}{\bibfnamefont{I.}~\bibnamefont{Bar-Joseph}},
  \bibinfo{journal}{Appl. Phys. Lett.} \textbf{\bibinfo{volume}{88}},
  \bibinfo{pages}{264103} (\bibinfo{year}{2006}).

\bibitem[{\citenamefont{Higbie et~al.}(2005)\citenamefont{Higbie, Sadler,
  Inouye, Chikkatur, Leslie, Moore, Savalli, and Stamper-Kurn}}]{Hig2005}
\bibinfo{author}{\bibfnamefont{J.~M.} \bibnamefont{Higbie}},
  \bibinfo{author}{\bibfnamefont{L.~E.} \bibnamefont{Sadler}},
  \bibinfo{author}{\bibfnamefont{S.}~\bibnamefont{Inouye}},
  \bibinfo{author}{\bibfnamefont{A.~P.} \bibnamefont{Chikkatur}},
  \bibinfo{author}{\bibfnamefont{S.~R.} \bibnamefont{Leslie}},
  \bibinfo{author}{\bibfnamefont{K.~L.} \bibnamefont{Moore}},
  \bibinfo{author}{\bibfnamefont{V.}~\bibnamefont{Savalli}}, \bibnamefont{and}
  \bibinfo{author}{\bibfnamefont{D.}~\bibnamefont{Stamper-Kurn}},
  \bibinfo{journal}{Phys. Rev. Lett.} \textbf{\bibinfo{volume}{95}},
  \bibinfo{pages}{0504011} (\bibinfo{year}{2005}).

\bibitem[{\citenamefont{Vengalattore et~al.}(2006)\citenamefont{Vengalattore,
  Higbie, Sadler, Leslie, and Stamper-Kurn}}]{Ven2006}
\bibinfo{author}{\bibfnamefont{M.}~\bibnamefont{Vengalattore}},
  \bibinfo{author}{\bibfnamefont{J.~M.} \bibnamefont{Higbie}},
  \bibinfo{author}{\bibfnamefont{L.~E.} \bibnamefont{Sadler}},
  \bibinfo{author}{\bibfnamefont{S.~R.} \bibnamefont{Leslie}},
  \bibnamefont{and}
  \bibinfo{author}{\bibfnamefont{D.}~\bibnamefont{Stamper-Kurn}}
  (\bibinfo{year}{2006}), \bibinfo{note}{to be submitted}.

\end{thebibliography}

\end{document}